\documentclass[preprint,aps]{revtex4-1}

\usepackage{graphicx}
\usepackage{amsmath}

\usepackage{pdfpages}
\usepackage{placeins}

\renewcommand{\rho}{\varrho}
\renewcommand{\phi}{\varphi}

\newcommand{\del}{\partial}
\newcommand{\ab}{\alpha \beta}

\begin{document}

\title{Border forces and friction control epithelial closure dynamics}

\author{Olivier Cochet-Escartin, Jonas Ranft, Pascal Silberzan and Philippe Marcq}
\affiliation{
Physico-Chimie Curie, 
Institut Curie, CNRS, Universit\'e Pierre et Marie Curie,
26 rue d'Ulm, F-75248 Paris Cedex 05 France
}

\date{March 5th, 2013}


\maketitle

\textbf{
Epithelization, the process whereby an epithelium covers a cell-free surface, 
is not only central to wound healing \cite{Sonnemann2011}
but also pivotal in embryonic morphogenesis \cite{Wood2002}, 
regeneration, and cancer \cite{Friedl2009}. 
In the context of wound healing, the epithelization mechanisms differ 
depending on the sizes and geometries of the wounds as well as 
on the cell type 
\cite{Brock1996,Grasso2007,Tamada2007,Abreu-Blanco2012,Anon2012}, 
while a unified theoretical decription is still lacking
\cite{Almeida2009,Arciero2011,Lee2011}. 
Here, we used a barrier-based protocol \cite{Poujade2007} that allows 
for making large arrays of well-controlled circular model wounds within 
an epithelium at confluence, without injuring any cells. 
We propose a physical model that takes into account border forces,
friction with the substrate, and 
tissue rheology. Despite the presence of a contractile actomyosin cable 
at the periphery of the wound, epithelization was mostly driven by border 
protrusive activity. Closure dynamics was quantified by an epithelization 
coefficient $D = \sigma_p/\xi$ defined as the ratio of the border protrusive 
stress $\sigma_p$ to the friction coefficient $\xi$
between epithelium and substrate. The same assay and model showed a high 
sensitivity to the RasV12 mutation on human epithelial cells, 
demonstrating the general applicability of the approach and its potential 
to quantitatively characterize metastatic transformations.
}

The experiments relied on the constraint of epithelial growth 
by cylindrical polydimethylsiloxane (PDMS) pillars whose base,
of radius $R_{\mathrm{w}}$, was in contact with the surface of 
a glass coverslip, therefore preventing cell growth on these areas 
(Fig.~\ref{fig:exp}, Supplementary Fig.~5). 
Removing the pillars (``barriers'') yielded a continuous cell monolayer 
surrounding circular cell-free patches (``wounds''), whose surface
did not differ from the rest of the substrate. 
Pillar removal did not injure the cells but triggered their migration
into the cell-free regions \cite{Anon2012,Poujade2007}. 
The free surface area $S(t)$ of each wound was dynamically monitored 
and we defined an effective radius $R(t)=\sqrt{S(t)/\pi}$, 
from which the margin velocity 
was computed. Experiments were first carried out on the well-known 
Madin-Darby canine kidney (MDCK) cells that are prototypical of a 
cohesive epithelium \cite{Gaush1966}. 
We then studied the influence 
of an oncogenic transformation on epithelization for the
human embryonic kidney line \cite{Ayllon2000}.

MDCK circular wounds of initial radii 
between $100 \, \mu$m and $250 \, \mu$m
rapidly lost their circular shape through the formation of 
leader cells at several positions around the margin (Supplementary Movie~1). 
Subsequently, these leader cells dragged multi-cellular fingers 
\cite{Omelchenko2003,Reffay2011} that eventually merged within the wound, 
thereby creating smaller secondary holes. These holes then proceeded to close, 
this time without leader cells or roughening of the margin 
(Supplementary Movie~2). When $R_{\mathrm{w}} > 100 \, \mu$m, 
this assay is qualitatively identical with barrier assays performed 
on large wounds of rectangular shape \cite{Poujade2007}.

In contrast, smaller wounds ($R_{\mathrm{w}} \le 100 \, \mu$m) healed 
without the formation of leader cells with only minor distortions 
of their disk-like shape (Fig.~\ref{fig:exp}, Supplementary Fig.~6). 
Of note, this transition radius 
is of the same order as the epithelial velocity correlation length 
\cite{Petitjean2010}. The closure of smaller wounds was completed 
within $20$ hours, and presented two striking features 
(Fig.~\ref{fig:exp}C-D).
\emph{(i)} Confocal imaging of F-actin revealed lamellipodia in variable number 
and size at the margin (Supplementary Movie~3). Active protrusions were not 
limited to the free surface of the wounds and we also systematically observed 
cryptic lamellipodia \cite{Farooqui2005} within the tissue
that did not show a preferential orientation (Supplementary Movie~6). 
\emph{(ii)} A pluricellular actomyosin cable was assembled at the margin 
only minutes after removal of the pillars and ran continuously around 
the wound edge.

The contribution of the actomyosin cable was assessed with  
2-photon laser ablation experiments (Supplementary Materials and Methods). 
Local ablations induced a retraction of the severed ends of the cable 
(Supplementary Fig.~8), on a time scale of a few seconds. 
This observation confirmed that the cable was under tension, as expected from 
the co-localization of F-actin in the cable with its associated molecular motor 
myosin II. Furthermore, we observed a small 
backward displacement ($1-2 \, \mu$m) of the edge of the wound in the radial 
direction upon ablation of the entire cable, on a time scale of a few minutes
(Supplementary Fig.~9, Supplementary Movie~7). Together, 
these results show that the cable exerts centripetal forces on the closing 
epithelium. 

To substantiate a physical model of epithelization, 
we used selective inhibition  to uncouple the contributions made 
by the contractile cable and by protrusive activity (Fig.~\ref{fig:inhib}).
Actomyosin contractility and lamellipodial activity are respectively 
associated with the small G-proteins Rho and Rac1 \cite{Ridley2001}.
Whereas the Rho inhibitor c3-transferase had little influence on the 
closure times  (Rho$^-$ assays, Supplementary Fig.~12A), 
the Rac1 inhibitor NSC-23766 induced a significant slowing down 
of the closing process (Rac$^-$ assays, Fig.~\ref{fig:inhib}A, 
Supplementary Fig.~12B).
Some Rac$^-$ MDCK wounds of large enough initial radii
did not close \cite{Farooqui2005}, and epithelization stopped at a
final nonzero value $R_e$ of the radius 
(Fig.~\ref{fig:inhib}B, Supplementary Movie~8). 
We conclude from these results that 
lamellipodial activity is the dominant driving force of epithelization 
\cite{Fenteany2000,Anon2012}.

Using velocimetry techniques \cite{Petitjean2010,Deforet2012},
we measured the velocity field around circular wounds in space and time. 
Strikingly, the angle-averaged radial component of the velocity 
decayed as the inverse of the distance $r$ to the initial center of the wound
(Supplementary Fig.~10), a signature of monolayer incompressibility 
(Supplementary Model). Indeed, the cell density was approximately uniform, 
and increased by less than $10 \, \%$ during closure,
since cells divided little or not at all (Supplementary Fig.~11).

On the basis of these observations, we model the tissue as a 
two-dimensional, isotropic, continuous material, 
whose flow is incompressible and driven by border forces (Supplementary Model).
The epithelium occupies at time $t$ the space outside a disk of radius $R(t)$,
with an initial radius $R_0 = R(t = 0)$ (Fig.~\ref{fig:model}A). 
We assume that lamellipodia exert a constant protrusive stress $\sigma_p$ 
at the margin, and that the friction force between epithelium and substrate 
is fluid, with a friction coefficient $\xi$. The radial force balance 
equation is integrated with a boundary condition at a cut-off radius
$r = R_{\mathrm{max}}$, a parameter of the model. A differential equation 
for $R(t)$ follows from the stress boundary condition at the border. 
Neglecting the contribution of the peripheral cable 
to force generation, and using an \emph{inviscid} tissue rheology, 
we obtain an analytical expression for the closure time 
$t_{\mathrm c} = t(R = 0)$ as a function of the initial radius $R_0$:
\begin{equation}
  \label{eq:closure}
  t_{\mathrm c}(R_0) = \frac{R_0^2}{4 D} \, \left( 1 + 
2 \ln\left( \frac{R_{\mathrm{max}}}{R_0} \right) \right),
\end{equation}
where the epithelization coefficient
$D = \sigma_p/\xi$ has the dimension of a diffusion coefficient.

Since the closure time is a robust quantity that depends little on the 
specifics of image analysis, we used equation~(\ref{eq:closure}) to fit
the data and measure the parameters $\sigma_p/\xi$ and $R_{\mathrm{max}}$
(Fig.~\ref{fig:tc}, Supplementary Fig.~13). 
We checked that taking into account force generation by the actomyosin 
cable in the stress boundary condition does not modify our results
(Supplementary Data Analysis, Supplementary Fig.~14A). 
We found that the cut-off radius $R_{\mathrm{max}}$, of the order of 
$110 \, \mu$m, varied little between different conditions. 
Compared to its wild type value 
($D_{\mathrm{wt}} = 353 \pm 38 \, \mu \mathrm{m}^2 \, \mathrm{h}^{-1}$,
$N = 130$), the epithelization coefficient was strongly reduced by Rac 
inhibition 
($D_{\mathrm{Rac}} = 198 \pm 22 \, \mu \mathrm{m}^2 \, \mathrm{h}^{-1}$, 
$N = 34$), 
and adopted an intermediate value under Rho inhibition 
($D_{\mathrm{Rho}} = 278 \pm 40 \, \mu \mathrm{m}^2 \, \mathrm{h}^{-1}$, 
$N = 30$).
Individual trajectories of wound radii were also satisfactorily 
fitted by the predicted time evolution of the radius $R(t)$
(Supplementary Equation~(13)),
and yielded estimates of the epithelization coefficient consistent within 
error bar with those obtained from closure time data, albeit with larger 
uncertainties (Supplementary Fig.~16). 
Since Rac inhibition impairs actin polymerization at the leading edge 
of migrating cells \cite{Ridley2001},
one expects a lower protrusive stress in Rac$^-$ assays,
conducive to a lower value of $\sigma_p/\xi$.
Both Rac and Rho inhibition may also modify the friction 
coefficient $\xi$, which generally depends on the intensity
and the dynamics of cell-substrate adhesion. This may explain 
the lower value of $\sigma_p/\xi$ measured under Rho inhibition
(Supplementary Data Analysis).

In order to check whether our results were robust against
varying assumptions on the epithelial rheology, we investigated 
the predicted closure dynamics of: \emph{(i)} a viscous epithelium, 
with a shear viscosity coefficient $\eta$;
and \emph{(ii)} an elastic epithelium, with a shear elastic modulus $\mu$.
Fitting data with the more complex functional forms of $t_{\mathrm c}(R_0)$  
thus obtained (Supplementary Model and Fig.~\ref{fig:model}D-E), 
we concluded that \emph{(i)} $\xi R_0^2 / \eta \gg 1$: external friction 
dominates internal viscosity \cite{Bonnet2012}; and \emph{(ii)} 
$\sigma_{\rm p}/\mu \gg 1$: protrusive forces dominate elastic forces
(Supplementary Data Analysis).
These results confirm that equation~(\ref{eq:closure}) 
provided a satisfactory description of the data on closing wounds.
Further, the trajectories of non-closing Rac$^-$ wounds 
could be fitted with the analytical expressions obtained on the basis
of an elastic epithelial rheology (Supplementary Fig.~17).
Due to Rac inhibition, the border force was small enough to allow 
a restoring elastic force to stop epithelization on the 
time scale of the experiment.

Finally, to test the sensitivity of the proposed quantification to 
cell phenotypes, we studied and compared epithelization by human embryonic 
kidney (HEK-HT) cells and by the derived cell line constitutively 
expressing the H-Ras oncogene (HEK-RasV12), using
the same experimental and data analysis protocols.
The dynamics were globally faster than what had been observed for MDCK cells 
(compare Supplementary Movies 3 and 4 or Figs.~\ref{fig:tc}A 
and \ref{fig:tc}B). Moreover, the HEK-RasV12 cell line had 
a greater protrusive activity than the HEK-HT line 
(compare Supplementary Movies 4 and 5). The model in its simplest form, 
equation~(\ref{eq:closure}), accounted well for the closure time data 
(Fig.~\ref{fig:tc}B). Further, HEK-RasV12 wounds were characterized 
by a larger  epithelization coefficient 
($D_{\mathrm{HEK-RasV}12} = 1531 \pm 363 \, \mu \mathrm{m}^2 \, \mathrm{h}^{-1}$, 
$N = 65$)
than HEK-HT wounds 
($D_{\mathrm{HEK-HT}} = 572 \pm 57 \, \mu \mathrm{m}^2 \, \mathrm{h}^{-1}$, 
$N = 63$).
The mutation carried by the HEK-RasV12 cell line 
is known to be common in different types of cancer \cite{Chin1999} 
and to promote angiogenesis \cite{Mali2010}
and cell motility \cite{Meadows2004}. 
The larger value of the epithelization coefficient for HEK-RasV12 than for 
HEK-HT wounds proves to be a signature of the metastatic capacity of 
the transformed cell line.

To summarize, a model of the epithelium as an inviscid fluid allowed to
quantify the closure of small circular wounds and to classify different 
cell phenotypes according to the value of the epithelization coefficient. 
The protrusive force  generated by lamellipodia at and close to the margin
drove collective migration. 
From the order of magnitude of the  epithelization coefficient 
$\sigma_p/\xi \approx 10^2 \, \mu \mathrm{m}^2 \, \mathrm{h}^{-1}$, and given
that of cellular  protrusive forces $F_p \approx 1$ nN \cite{Prass2006},
we deduce an order of magnitude of the epithelium-substrate friction 
coefficient $\xi \approx 1 \, \mathrm{nN} \, \mu\mathrm{m}^{-3} \, \mathrm{s}$
on a glass substrate (Supplementary Data Analysis). 
Down- or up-regulating integrin expression or turn-over 
may modify $\xi$, and in turn alter epithelization dynamics.  
Recent work has shown that the competition between 
friction  and flow  governs collective migration in developing
organisms \cite{Bonnet2012,Behrndt2012,Mayer2010}.
Appropriate modifications of the model may 
lead to quantitative descriptions of  \emph{in vivo} epithelization 
during wound healing \cite{Brock1996,Wood2002,Abreu-Blanco2012}, but also during
embryonic morphogenesis, as in, \emph{e.g.}, 
the dorsal closure of \emph{D. melanogaster} \cite{Kiehart2000}
or the ventral enclosure of \emph{C. elegans} \cite{Williams-Masson1997}.

\begin{acknowledgments}
The authors thank Olivier Leroy and Olivier Renaud, along with other members 
of the PICT-IBISA platform, for their help with confocal imaging and 
laser ablation experiments, as well as Maria-Carla Parrini for her help with 
immunostaining experiments. O.~C. acknowledges support by the 
Association pour la Recherche contre le Cancer.
\end{acknowledgments}


\newpage

\begin{figure}[!h]
\caption{\label{fig:exp} 
\textbf{Epithelization of small circular wounds} (wild-type MDCK cells)
\\
\textbf{A}: Field of view ($R_{\mathrm{w}} = 50 \, \mu$m). 
Between two and four such fields are recorded in a typical experiment.
Several adjacent MDCK wounds are visible at $t=0$ (left) 
and $t=3$ h (right) after removal 
of the PDMS pillars. Note the intrinsic diversity of closure dynamics.
The typical cell size is of the order of $15 \, \mu$m. 
\\
\textbf{B}: Timelapse zoomed on a single wound 
($R_{\mathrm{w}} = 37.5 \, \mu$m).\\ 
\textbf{C}: Wound fixed at $t=30$ min ($R_{\mathrm{w}} = 25 \, \mu$m)
and stained for phospho-myosin II light chain (red), 
F-actin (green) and nuclei (blue) by 
immunofluoresence. Note the presence of a pluricellular actomyosin cable 
and of lamellipodia (indicated by stars).  For this size, we observed 
between $0$ and $2$ lamellipodia whose area ranged between 
$20 \, \mu \mathrm{m}^2$ and $175 \, \mu \mathrm{m}^2$ ($N = 10$). 
Scale bars: $20 \, \mu$m.
\\
\textbf{D}: Section of a live wound (MDCK-LifeAct-GFP, 
$R_{\mathrm{w}} = 25 \, \mu$m, $t=30$ min) imaged by confocal microscopy. 
The position of the cable on both sides is indicated by arrows. 
Scale bar: $10 \, \mu$m.
}
\end{figure}

\begin{figure}[!h]
\caption{\label{fig:inhib} 
\textbf{Effect of Rac1-inhibitor on closure dynamics.}
\\
\textbf{A}: Closure time ($R_{\mathrm{w}} = 50 \, \mu$m).
Red: MDCK wild-type; blue: Rac$^-$ assay.
Box: first quartile, median and last quartile.
\\
\textbf{B}: Rac$^-$ assay, fraction of MDCK wounds proceeding to 
full closure within $18$~h for the initial sizes 
$R_{\mathrm{w}} = 25 \, \mu$m ($N=8$), 
$R_{\mathrm{w}} = 50 \, \mu$m ($N=39$) and 
$R_{\mathrm{w}} = 100 \, \mu$m ($N=16$).
}
\end{figure}

\begin{figure}[!h]
\caption{\label{fig:model} 
\textbf{Physical model of epithelial closure.}
\\
\textbf{A}: Sketch of a closing circular wound, of initial radius 
$R_0 = R(t = 0)$.
Two border forces may drive closure: $\sigma_p$ is the protrusive stress
produced by lamellipodia, $\gamma$ the line tension due to the contractile 
circumferential cable (see the stress boundary
condition Supplementary Eq.~(6))\\
\textbf{B-F}: Model predictions. Plots of the closure time $t_{\mathrm c}$ 
as a function of the initial effective radius $R_0$.\\
\textbf{B}: 
Effect of the variation of $D$ while 
$R_{\mathrm{max}} = 110 \, \mu \mathrm{m}$ is fixed,
inviscid rheology without cable, equation~(\ref{eq:closure}) 
(also Supplementary Eq.~(29)).\\
\textbf{C}: 
Effect of the variation of $R_{\gamma} = \gamma/\sigma_p$
while $D = 200 \, \mu \mathrm{m}^2 \, \mathrm{h}^{-1}$ and
$R_{\mathrm{max}} = 110 \, \mu \mathrm{m}$ are  fixed,
inviscid rheology with a cable, Supplementary Eq.~(30).\\
\textbf{D}: 
Effect of the variation of $R_{\eta} = \sqrt{\eta/\xi}$
while $D = 200 \, \mu \mathrm{m}^2 \, \mathrm{h}^{-1}$,
$R_{\mathrm{max}} = 110 \, \mu \mathrm{m}$ and 
$R_{\gamma} = 10 \, \mu \mathrm{m}$ are  fixed,
viscous rheology, Supplementary Eq.~(31).\\
\textbf{E}:
Effect of the variation of $\mu/\sigma_p$
while $D = 200 \, \mu \mathrm{m}^2 \, \mathrm{h}^{-1}$,
$R_{\mathrm{max}} = 110 \, \mu \mathrm{m}$ and 
$R_{\gamma} = 100 \, \mu \mathrm{m}$ are  fixed,
elastic rheology, Supplementary Eq.~(32).
When $\mu/\sigma_p = 1$, closure is complete and characterized by a finite
closure time only below a value of $R_0$ 
above which elastic forces are strong enough to stop epithelization.
}
\end{figure}

\begin{figure}[!h]
\caption{\label{fig:tc} 
\textbf{Physical parameters of epithelization.}
\\
\textbf{A-B}: Closure time $t_{\mathrm c}$ (filled circles) as a function 
of the initial effective radius $R_0$, fitted by equation~(\ref{eq:closure}) 
(solid curves) with the constraints $D, R_{\mathrm{max}} \ge 0$.
One circle corresponds to one wound.
\\
\textbf{A}: MDCK wounds. 
Wild Type ($D = 353 \pm 38 \, \mu \mathrm{m}^2 \, \mathrm{h}^{-1}$,
$R_{\mathrm{max}} = 117 \pm 11 \, \mu \mathrm{m}$, $N = 130$), 
Rho$^-$ assay ($D = 278 \pm 40 \, \mu \mathrm{m}^2 \, \mathrm{h}^{-1}$,
$R_{\mathrm{max}} = 114 \pm 14 \, \mu \mathrm{m}$, $N = 30$) and  
Rac$^-$ assay 
($\sigma_p/\xi = 198 \pm 22 \, \mu \mathrm{m}^2 \, \mathrm{h}^{-1}$,
$R_{\mathrm{max}} = 105 \pm 9 \, \mu \mathrm{m}$, $N = 34$).\\
\textbf{B}: HEK-HT assay
($\sigma_p/\xi = 572 \pm 57 \, \mu \mathrm{m}^2 \, \mathrm{h}^{-1}$,
$R_{\mathrm{max}} = 132 \pm 12 \, \mu \mathrm{m}$, $N = 63$) 
and HEK-RasV12 assay
($\sigma_p/\xi = 1531 \pm 363 \, \mu \mathrm{m}^2 \, \mathrm{h}^{-1}$,
$R_{\mathrm{max}} = 223 \pm 77 \, \mu \mathrm{m}$, $N = 65$). 
\\
\textbf{C-F}:
Epithelization coefficient $D$ and cut-off radius $R_{\mathrm{max}}$.
\textbf{C}, \textbf{D}: MDCK wounds.
\textbf{E}, \textbf{F}: HEK wounds.
Error bars correspond to $95 \, \%$ confidence level.
}
\end{figure}

\newpage
\centerline{\includegraphics[width=1.5\linewidth]{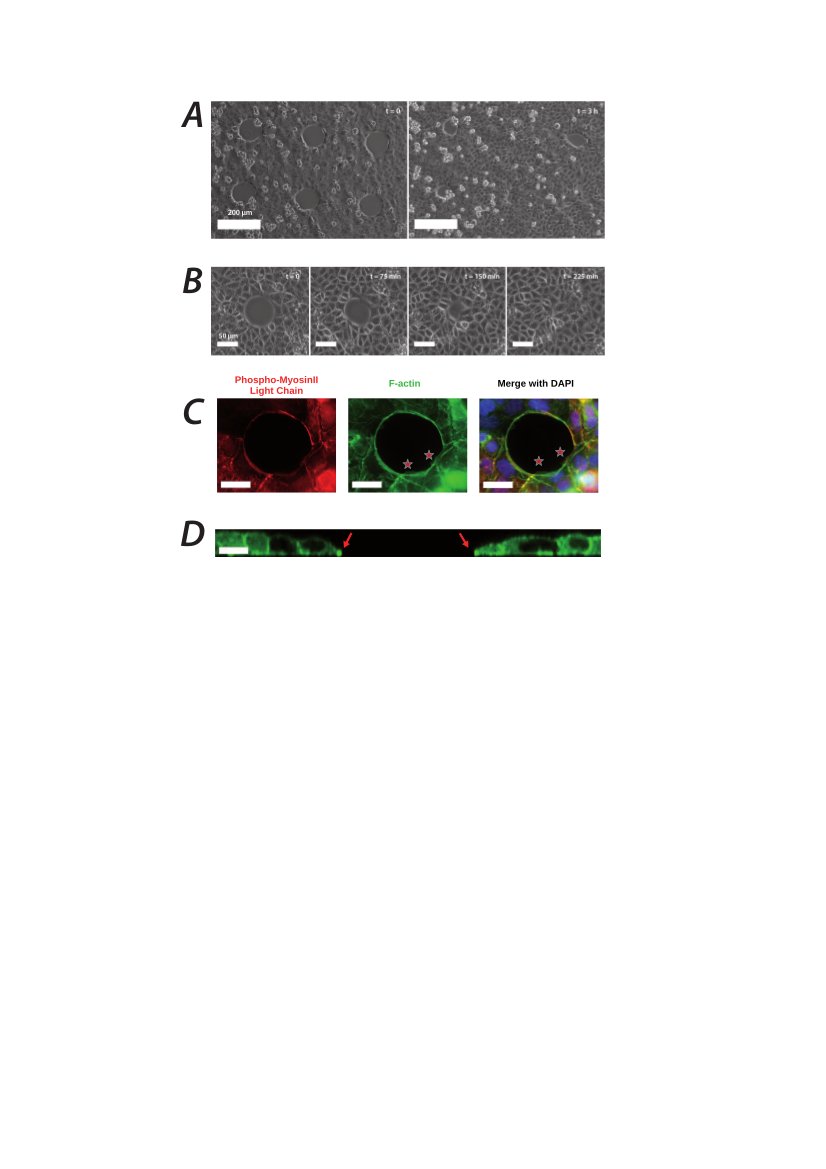}}

\newpage
\vspace*{-8.0cm}
\hspace*{-3.0cm}
\includegraphics[width=1.1\linewidth]{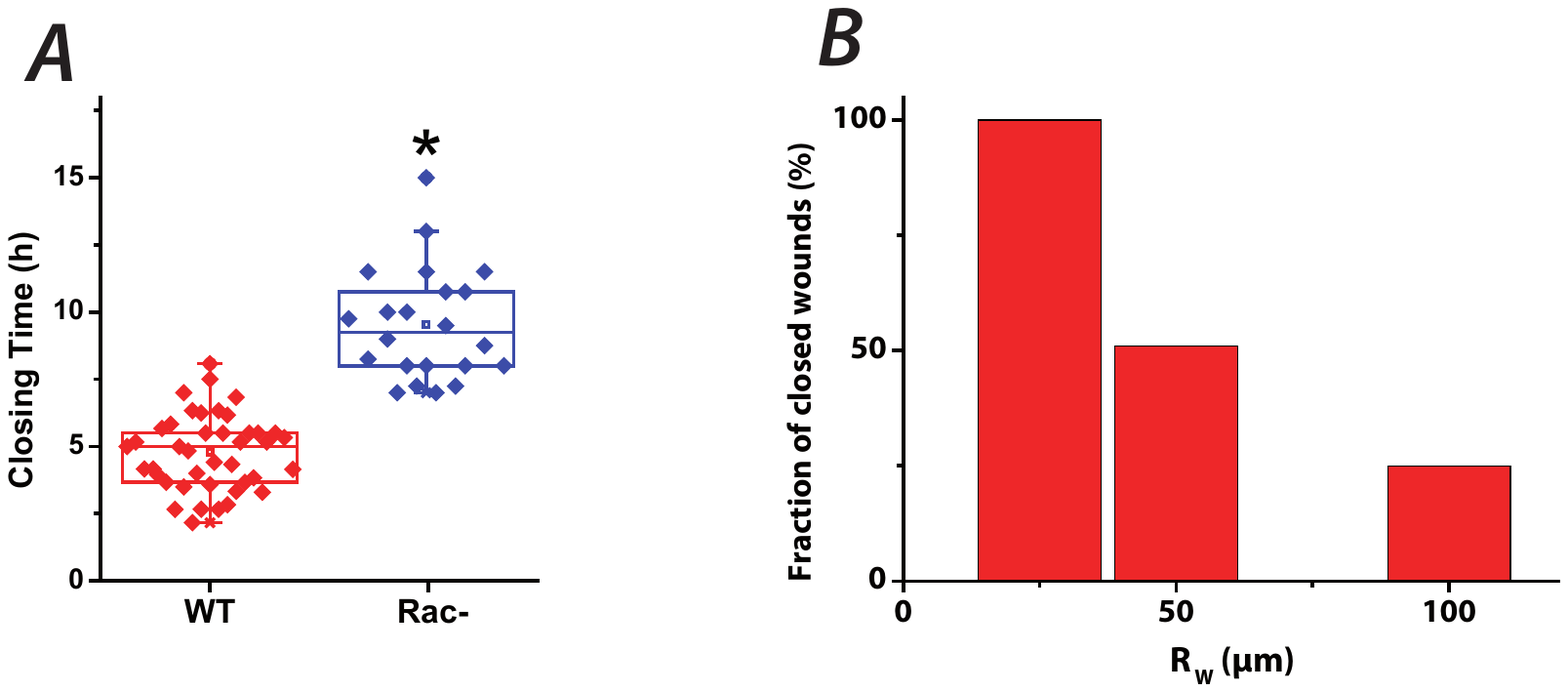}

\newpage
\textbf{A}
\centerline{\includegraphics[width=6.4cm]{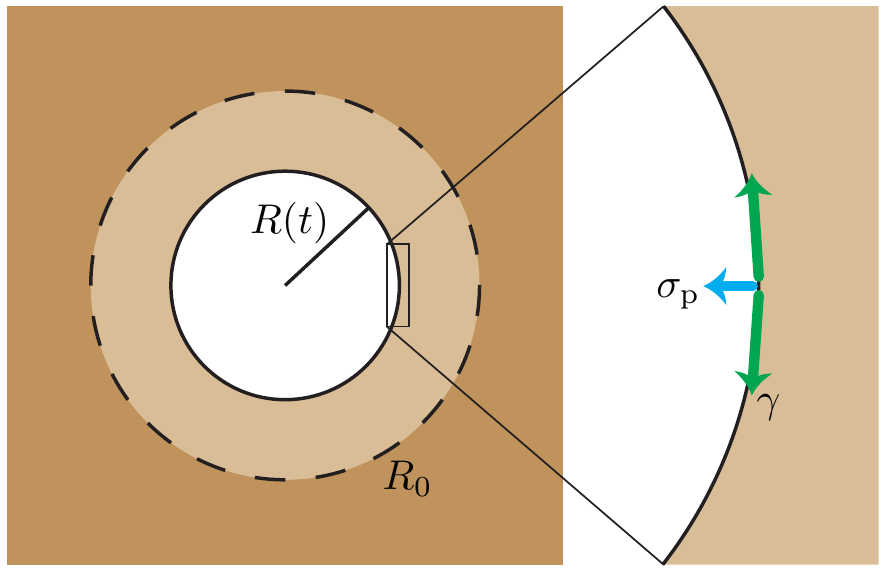}}\\
\textbf{B}
\includegraphics[width=7.7cm]{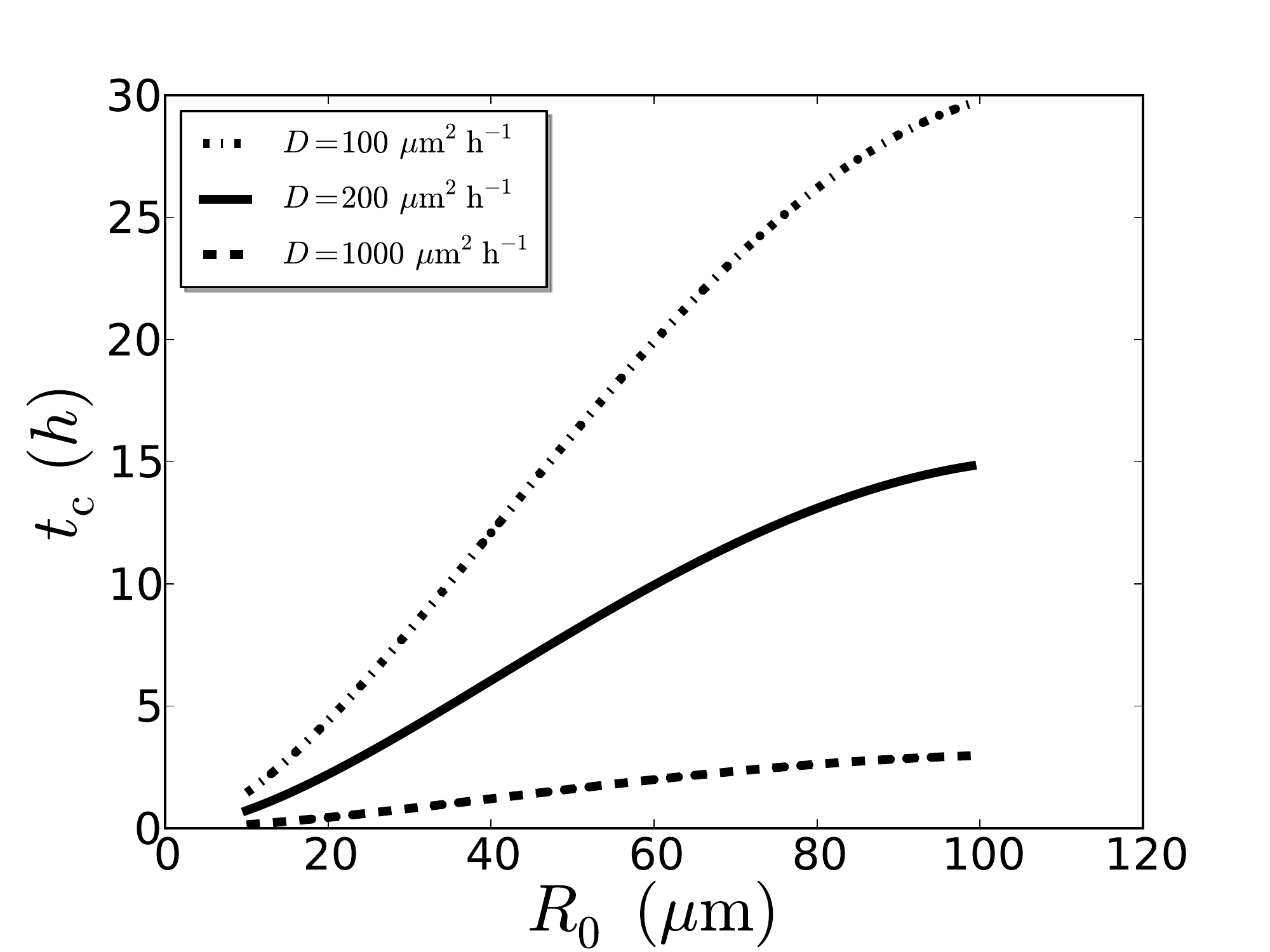}
\textbf{C}
\includegraphics[width=7.7cm]{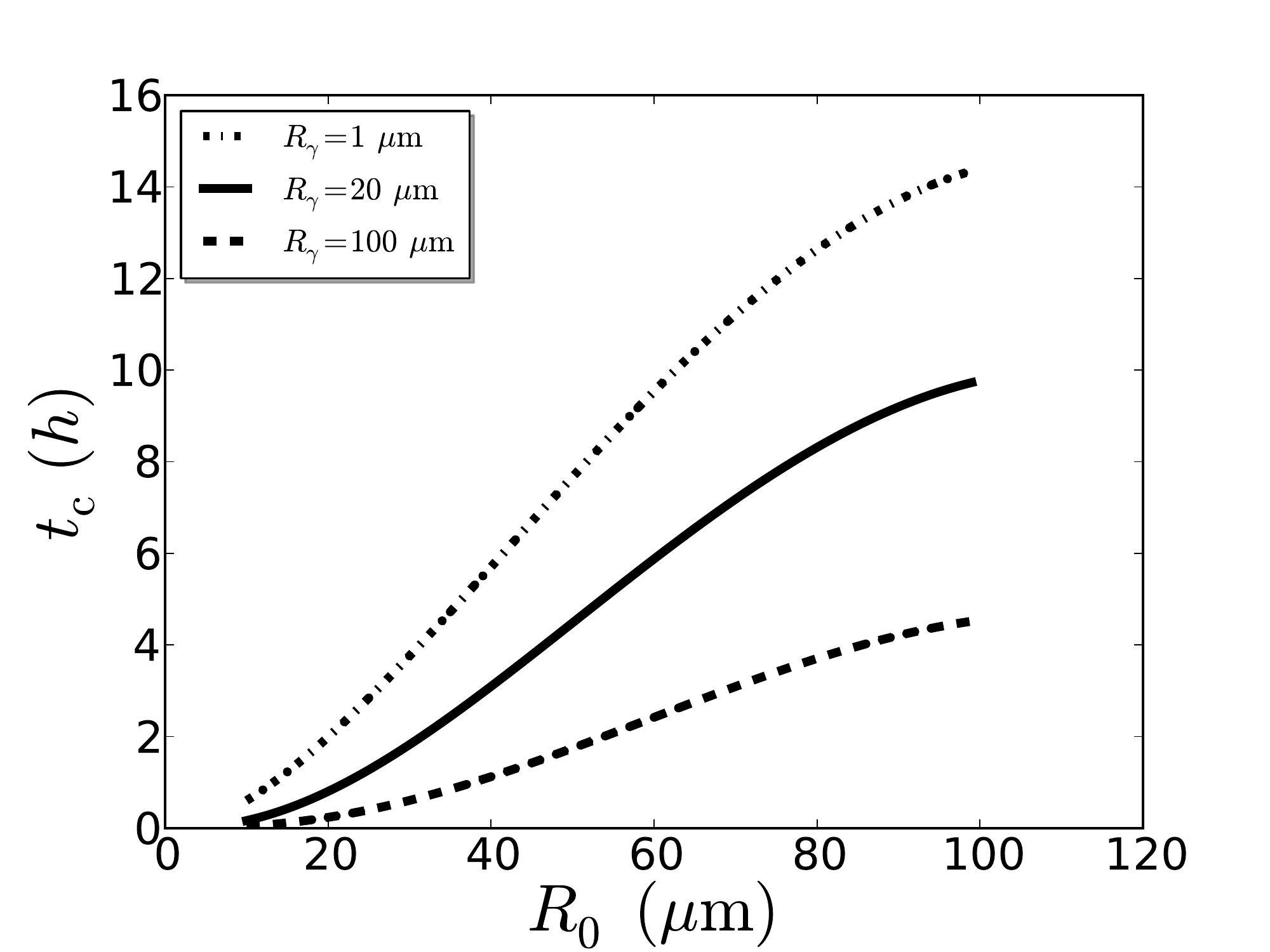}
\textbf{D}
\includegraphics[width=7.7cm]{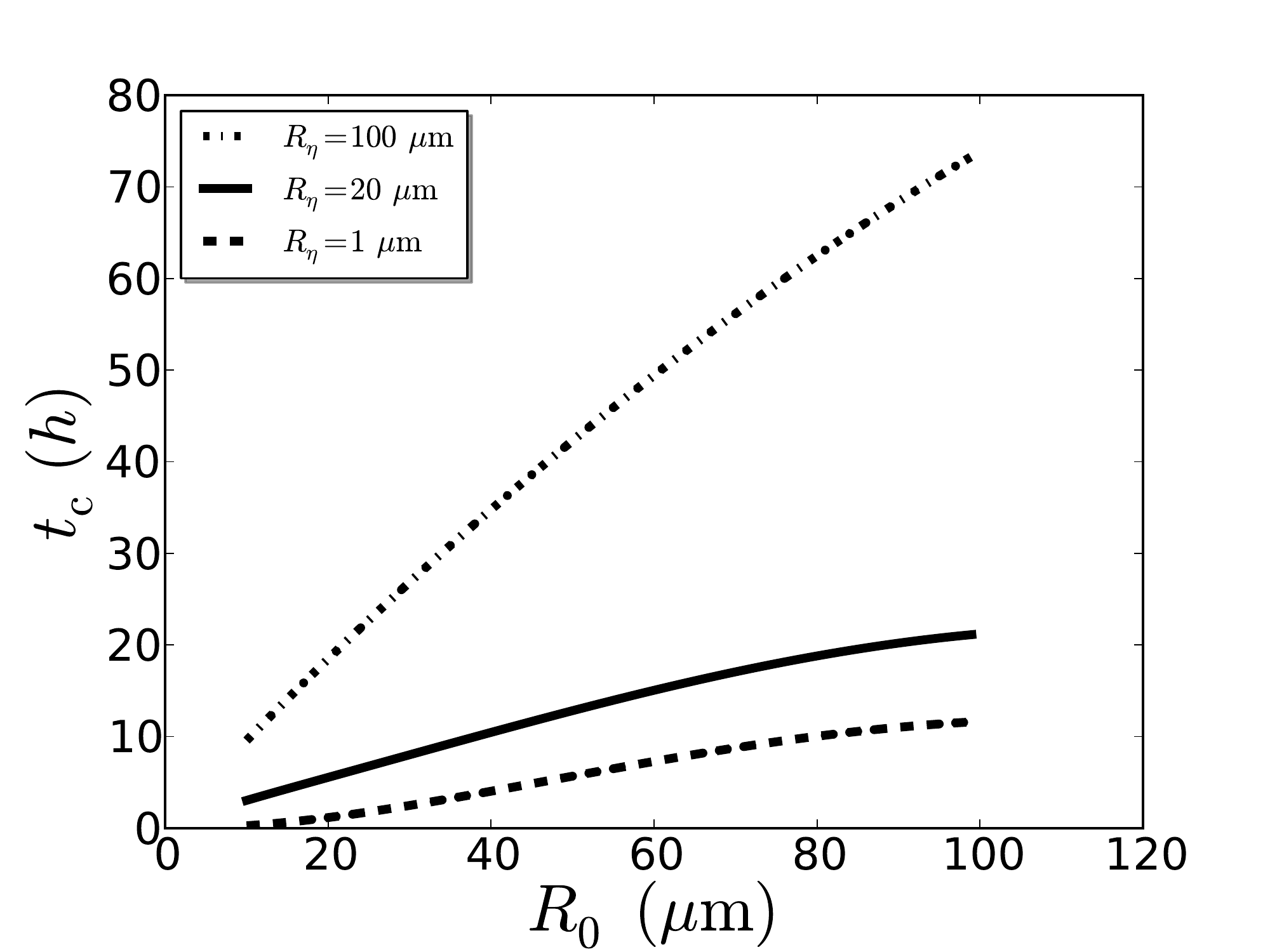}
\textbf{E}
\includegraphics[width=7.7cm]{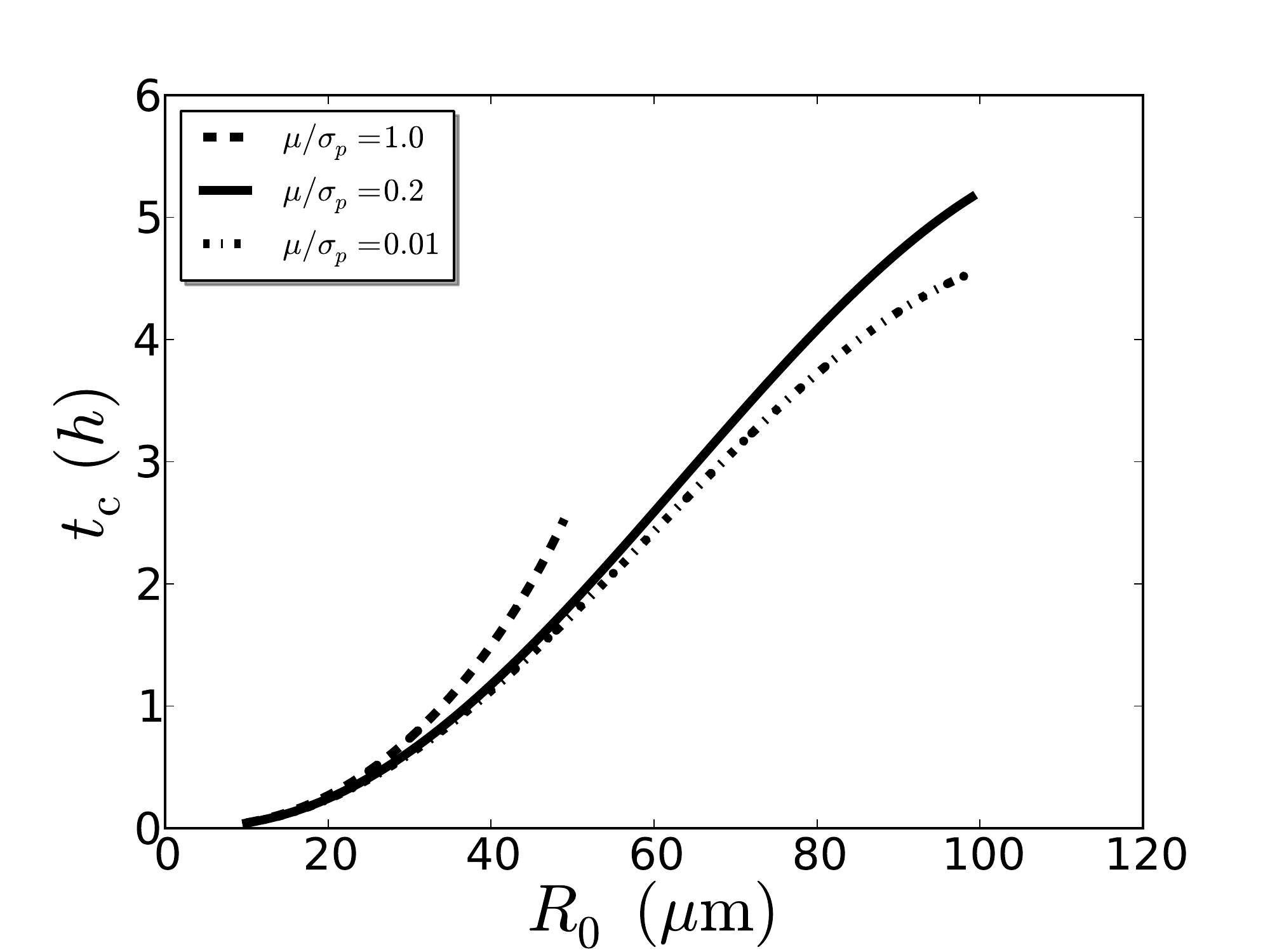}

\newpage
\includegraphics[width=\linewidth]{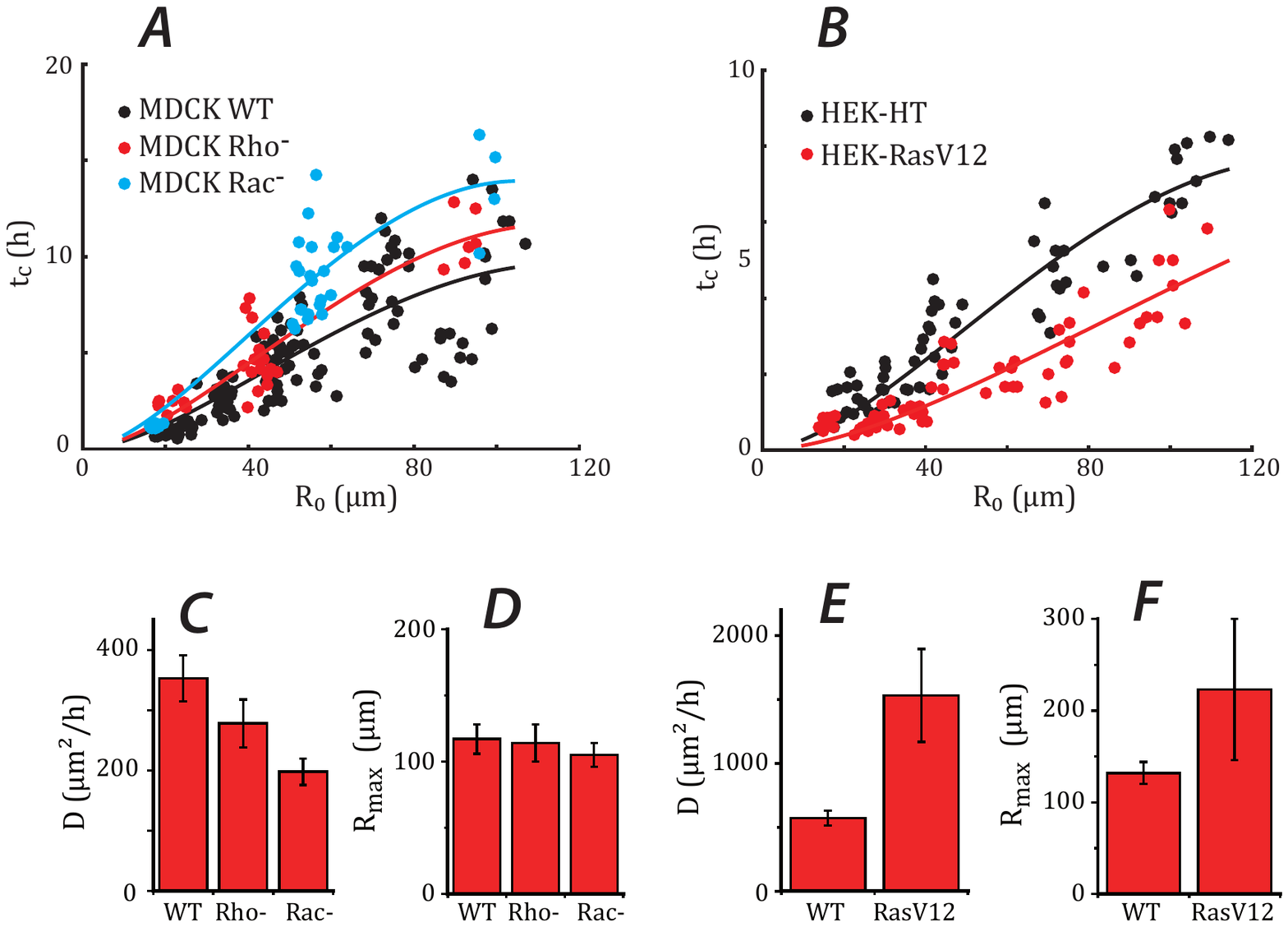}

\newpage
\setcounter{equation}{1}
\setcounter{figure}{4}
\setcounter{section}{0}

\centerline{\large \bf Supplementary Information}

\bigskip
\bigskip

\section{Materials and methods}
\label{sec:mat:meth}

\subsection{Cell culture}
\label{sec:mat:meth:cell}

MDCK wild type cells were cultured in Dulbecco's modified Eagle's medium 
(Gibco) supplemented with $10 \%$ FBS (Sigma), $2$ mM L-glutamin solution 
(Gibco) and $1 \%$ antibiotics solution [penicillin ($10000$ units/mL)  
+ streptomycin ($10$ mg/mL), Gibco] at $37^o$C, $5 \%$ CO$_2$ and 
$90 \%$ humidity. The LifeAct-GFP transfected cells were cultured 
in the same medium, supplemented with $400 \, \mu$g/mL geneticin (Invitrogen).
Other derived MDCK lines were used (histone-mCherry, cadherin-GFP, actin-GFP) 
and were cultured in the same way as the LifeAct-GFP line.

HEK-HT-wild type cells were cultured in Dulbecco's modified Eagle's medium 
(Gibco) supplemented with $10 \%$ FBS (Sigma), $2$ mM L-glutamin solution 
(Gibco), $1 \%$ antibiotics solution [penicillin ($10000$ units/mL)  + 
streptomycin ($10$ mg/mL), Gibco], $100 \, \mu$g/mL hygromycin solution 
(Life Technologies) and $400 \, \mu$g/mL geneticin (Invitrogen) 
at $37^o$C, $5 \%$ CO$_2$ and $90 \%$ humidity. The HEK-HT-RasV12 cells 
were cultured in the same medium, supplemented with $0.5 \, \mu$g/mL 
puromycin solution (Life Technologies).

The various inhibitors (Y-27632 (Sigma), C3-transferase (Tebu-bio) 
and NSC-23766 (Tocris)) were perfused in the flow chamber $2$h 
before removing the PDMS template. We used concentrations of $50 \, \mu$M 
for Y-27632, $1 \, \mu$g/mL for C3-transferase and $50\,\mu$M for NSC-23766.

\subsection{PDMS pillars preparation}
\label{sec:mat:meth:pillar}

The PolyDiMethylSiloxane (PDMS, Sylgard 184, Dow Corning) pillars 
were molded on a photoresist template obtained by classic lithography 
techniques. $100 \, \mu$m and $200 \, \mu$m-thick circular structures were 
fabricated in negative photoresist (SU8-2100, Microchem) by photolithography. 
Uncured PDMS was then poured on this template and $1$ mm spacers were used 
to constraint the height of the whole structure. It was then cured in a 
$65^o$C oven over night. With this technique, thousands of pillars of 
different radii can be manufactured at once. This PDMS stamp was then 
manually cut to the right dimension for each experiment.

\begin{figure}[!t]
\includegraphics[width=\linewidth]{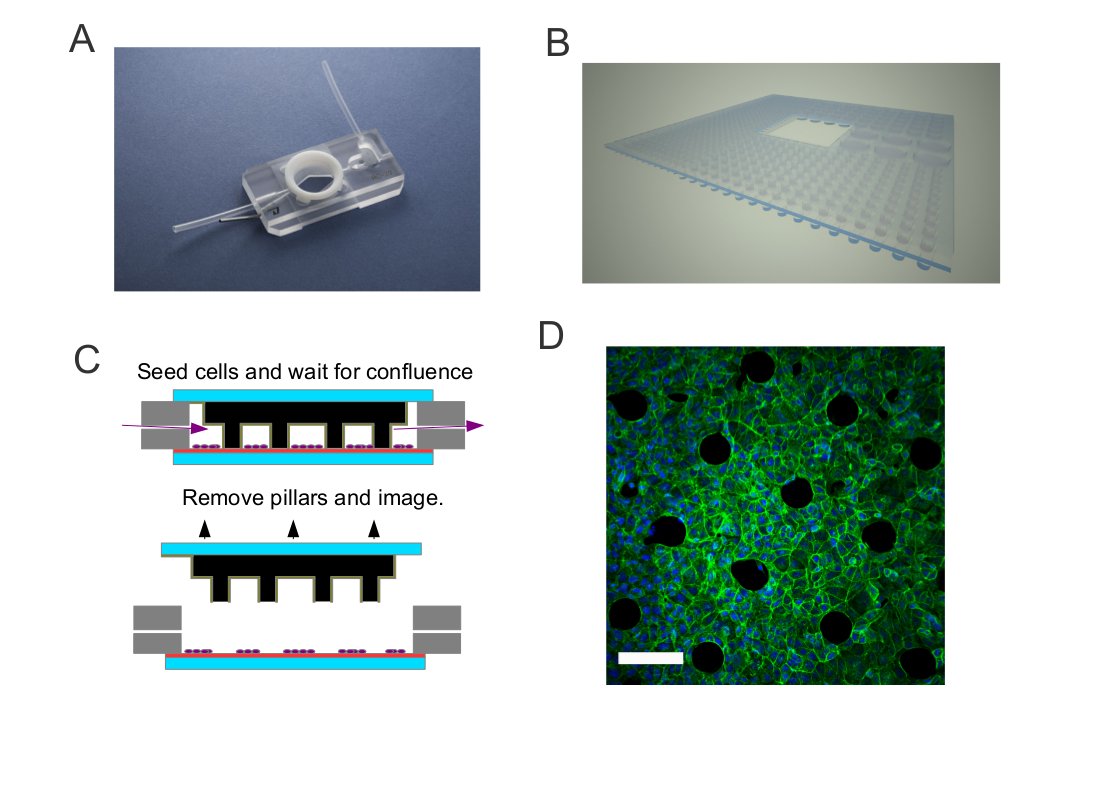}
\caption{\label{fig:protocol} 
\textbf{Experimental protocol and initial conditions.}\\
\textbf{A:} Picture of the flow chamber (Warner Instruments, model RC-20h).\\
\textbf{B:} Schematics of the PDMS template.\\ 
\textbf{C:} Schematics of the protocol. Cells are allowed to reach 
confluence before the template is removed.\\
\textbf{D:} MDCK wounds ($R_{\mathrm{w}} = 25 \, \mu$m) were fixed 
while constrained under the template and labeled
for F-actin (green) and nuclei (blue). Scale bar: $100 \, \mu$m.
}
\end{figure}

\subsection{Experimental protocol}
\label{sec:mat:meth:protocol}

Cells were cultured in a small volume flow chamber (Warner Instruments, 
RC20-h model, Fig.~\ref{fig:protocol}). The chamber is sealed on top and 
bottom by $\#1$ $15$ mm glass coverslips (Delta). The bottom coverslip was 
treated with $100 \, \mu$g/mL fibronectin (Life Technologies) in PBS at 
room temperature for one hour before being added to the chamber. 
The top coverslip was irreversibly bounded to the PDMS template by 
treating them both for $30$ s in an air plasma. They were then both 
treated with poly L-lysine-Polyethylene Glycol (PLL-PEG, Susos) at 
$0.1$ mg/mL for $5$ minutes to ensure that cells did not adhere to the 
pillars. The chamber was then hermetically sealed with silicone high 
vacuum grease. Cells were then seeded in the chamber at high concentrations 
($\approx 5 \, 10^4$ cells/$\mu$L) and allowed to adhere for one hour. 
Medium was then manually renewed every $30$~min to ensure proper growth 
underneath the template. Under these conditions, the cells reached 
confluence after $6$~h growth, the PDMS template was then delicately 
removed with the top coverslip and fresh medium was added to the chamber 
before imaging. Throughout this study, the initial time $t = 0$ corresponds by
convention to the time when the first image was acquired, unless 
explicitly mentioned otherwise.

\subsection{Image acquisition and treatment}
\label{sec:mat:meth:image}

The dynamics of closure were imaged in phase contrast on an Olympus IX-71 
inverted microscope equipped with thermal and atmospheric regulation (LIS). 
Images were acquired by a CCD-camera (Retiga 4000R, QImaging) and the setup 
was controlled by Metamorph (MetaImaging). The typical delay between two 
successive images was set between $3$ min and $15$ min depending on the 
initial sizes of the wounds and we used $10$x and $20$x objectives.
Confocal imaging of either live or fixed cells was performed under 
a LSM 710 NLO inverted confocal microscope (Zeiss) equipped with 
thermal and atmospheric regulation.
Images were then treated using ImageJ \cite{Note1}
and the free surface 
was computed through a masking algorithm based on a Fourier filter, 
an edge detection algorithm and, finally, binarization of the resulting image. 
This process proved robust. However, we checked by hand on several 
significative examples that the apparent distribution of initial radii 
(Fig.~4A) for one wound size resulted from the margin of 
error of this technique and from intrinsic variability 
and not from an actual distribution of initial 
radii that could have been due to variations in the microfabrication process.
The raw data on closure dynamics was then analysed with Matlab 
(Mathworks).

\begin{figure}[!t]
\includegraphics[width=0.6\linewidth]{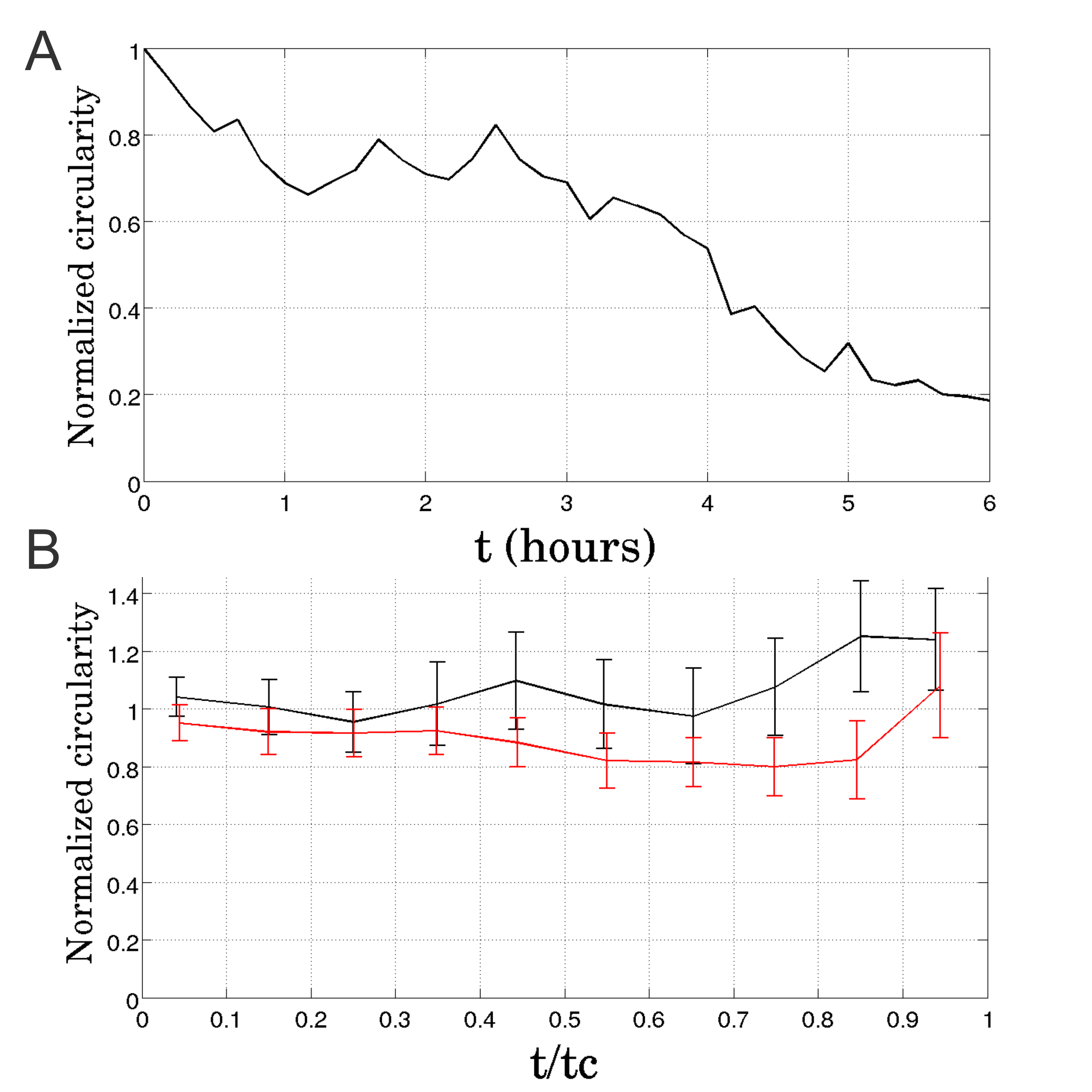}
\caption{\label{fig:circ} 
\textbf{Circularity dynamics.}\\
\textbf{A:}
Plot of the normalized circularity $c(t)/c(0)$ (Eq.~(\ref{eq:def:circ}))
\textrm{vs.} time $t$ of the large wound in Supplementary Movie 1
($R_{\mathrm{w}} = 250 \, \mu$m, red curve).
The measurement stops when the fingers merge at $t = 6$h.\\
\textbf{B:}
Plot of the normalized circularity $c(t)/c(0)$ 
\textrm{vs.} normalized time $t/t_c$, for the smallest 
($R_{\mathrm{w}} = 25 \, \mu$m, $N=18$, black curve) and the largest
initial radii ($R_{\mathrm{w}} = 100 \, \mu$m, $N=21$, red curve) 
of the small wounds. Error bars indicate the s.e.m. 
}
\end{figure}

\begin{figure}[!h]
\includegraphics[width=\linewidth]{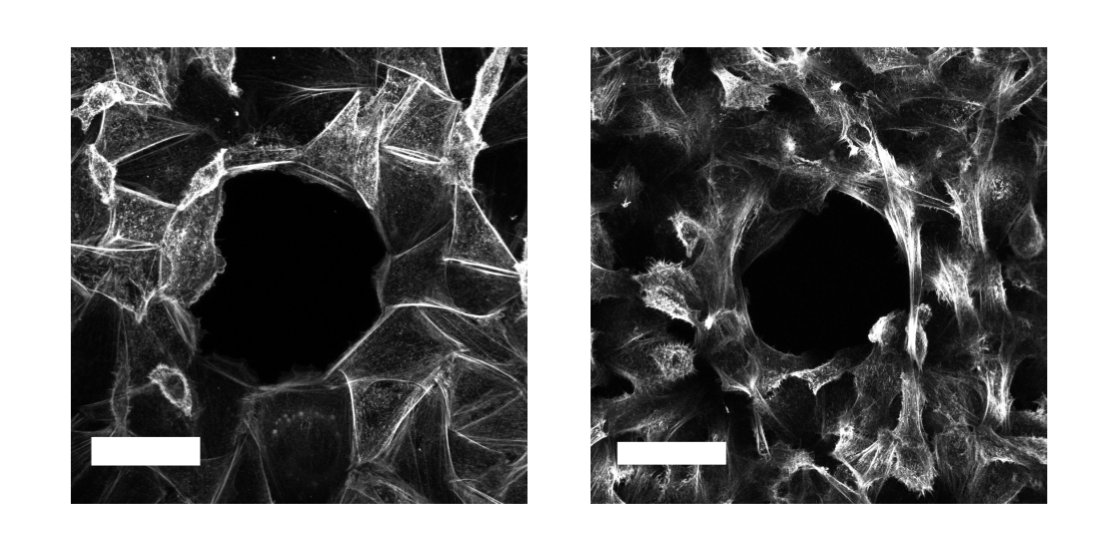}
\caption{\label{fig:HEK:staining} 
\textbf{Actin staining on HEK cells.}\\
HEK-HT (left) and HEK-RasV12 (right) wounds ($R_{\mathrm{w}} = 50 \, \mu$m)
were allowed to close for $30$ min and were then fixed and stained 
for F-actin with phalloidin. 
Numerous lamellipodia of different numbers and sizes are observed in both cases.
Scale bar: $50 \, \mu$m.
}
\end{figure}

\subsection{Circularity measurements}
\label{sec:mat:meth:circularity}

Wound shape was quantified by the circularity
\begin{equation}
  \label{eq:def:circ}
c(t) = \frac{4 \pi S(t)}{P(t)^2} \, ,  
\end{equation}
where $P(t)$ and $S(t)$ denote respectively the perimeter and 
the area of the wound at time $t$. 
This definition yields $c=1$ for a perfect circle, $c=0$ for a fractal 
structure with finite area but infinite perimeter, and 
in general $0 \le c \le 1$ for a closed curve. However, 
this measurement depends on the resolution of images,
due to pixelization artifacts \cite{Bottema2000}. We therefore normalized 
the circularity of each wound by its initial value $c(t = 0)$ 
(Fig.~\ref{fig:circ}).

\begin{figure}[!t]
\includegraphics[width=\linewidth]{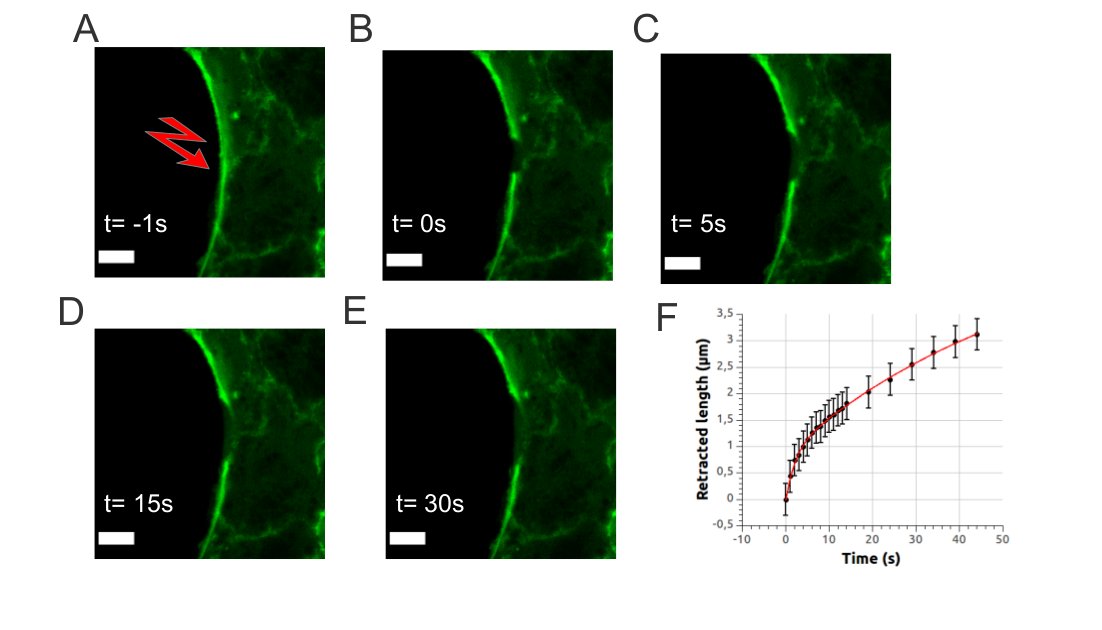}
\caption{\label{fig:ablation:local} 
\textbf{Local ablation of the acto-myosin cable.}\\
\textbf{A-E:} Timelapse of the retraction of an acto-myosin cable
after laser ablation (MDCK-LifeAct-GFP wound, $R_{\mathrm{w}} = 25 \, \mu$m),
imaged through confocal microscopy at $t=-1$~s, 
$0$~s, $5$~s, $15$~s and $30$~s. 
Here $t = 0$~s corresponds to the first image acquired immediately 
after ablation. Scale bar: $5 \, \mu$m.\\ 
\textbf{F:} Retraction dynamics of one of the severed ends of the cable
(black circles) with a double exponential fit (red curve)
$\Delta l(t) = l_1 \left( 1 - e^{-t/\tau_2} \right)
+ l_2 \left( 1 - e^{-t/\tau_2} \right)$. 
The fit yields two characteristic times,  $\tau_1 = 59.9 \pm 26.0$ s and 
$\tau_1 = 2.0 \pm 0.4$ s and two retraction lengths, 
$l_1 = 4.3 \pm 1.2 \, \mu$m and $l_2 = 0.9 \pm 0.1 \, \mu$m.
}
\end{figure}

\subsection{Immunofluorescent stainings}
\label{sec:mat:meth:staining}

Cells were fixed with $4 \%$ paraformaldehyde for $15$ min, permeabilized 
with $0.1 \%$ Triton-X100 for $10$ min, saturated in PBS supplemented 
with $10 \%$ FBS for $20$~min. Myosin labelling was performed by incubation 
for one hour with a rabbit anti-phospho Myosin Light Chain antibody (Ozyme) 
at $1$:$100$ before staining for one hour with a Cy-3 conjugated donkey 
anti-rabbit (Ozyme) used at $1$:$500$. Actin was stained with an 
alexa488-conjugated phalloidin (Life Technologies) at $1$:$1000$. Finally, 
the cells were mounted using Anti Fade Gold Reagent with DAPI 
(Life Technologies). The samples were then imaged on a 
LSM 710 NLO inverted confocal microsope (Zeiss), 
see Fig.~\ref{fig:HEK:staining}).

\begin{figure}[!t]
\includegraphics[width=\linewidth]{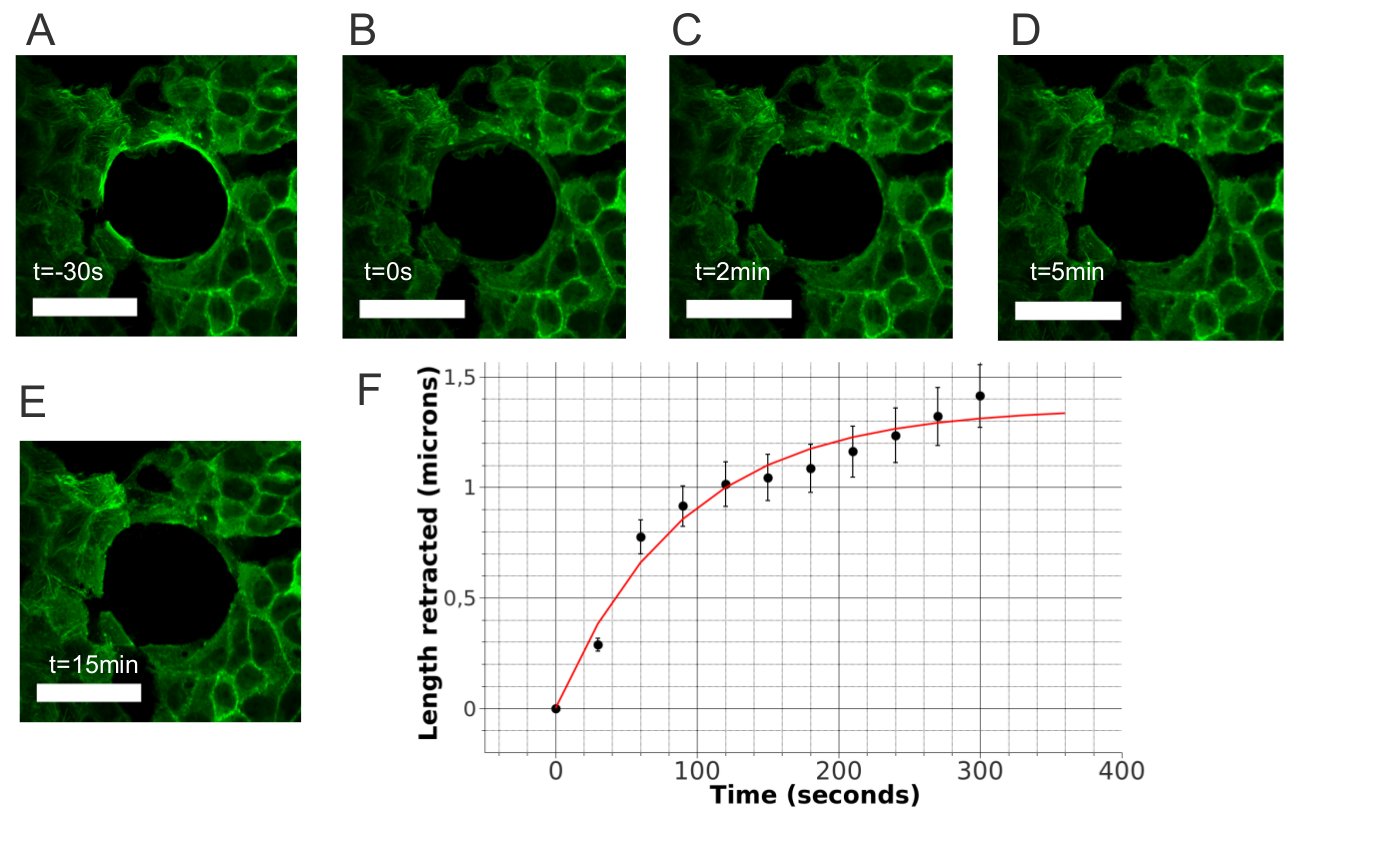}
\caption{\label{fig:ablation:global} 
\textbf{Ablation of the entire cable.}\\
\textbf{A-E:} Timelapse of the retraction of the wound edge after laser 
ablation of the circumferential cable (MDCK-LifeAct-GFP wound, 
$R_{\mathrm{w}} = 25 \, \mu$m), imaged through confocal microscopy at $t=-30$~s, 
$0$~s, $2$~min, $5$~min and $15$~min. Here $t = 0$~s corresponds to the 
first image acquired immediately after ablation. Scale bar: $50 \, \mu$m.
The actin cable was clearly apparent before ablation. \\ 
\textbf{F:} Plot of the retracted wound radius as a function of time 
(black circles), fitted by an exponentially decaying function of time
$\Delta R(t) = l \left( 1 - e^{-t/\tau} \right)$ 
(red curve). We obtain  a retracted length of 
$l = 1.36 \pm 0.15 \,\mu$m and a retraction time of $\tau = 90.7 \pm 27.9$~s.
}
\end{figure}

\subsection{Laser ablation}
\label{sec:mat:meth:ablation}

Laser ablation experiments were performed under a 
LSM 710 NLO (Zeiss) inverted confocal microscope  with a $64$x objective. 
The microscope was coupled to a femtosecond pulsed (pulse duration 
shorter than $100$~fs) 
2-photon Mai-Tai HP laser (Spectra Physics). For ablation, the wavelength 
and output power were respectively set at $810$~nm and around $0.1$~W. 
Between ten and twenty iterations of the ablation were applied to a 
zone drawn by hand through the Zen software (Zeiss) leading to a pixel 
dwell between $100 \,\mu$s and $200 \, \mu$s.

We first performed local ablations, and recorded the time course of the
retracted length of the cable (Fig.~\ref{fig:ablation:local}).
Rather than using one exponentially decaying function of time, fitting 
by the sum of two exponentially decaying functions provided better 
agreement with data. 
The two characteristic times, of the order of seconds and minutes respectively, 
differed by an order of magnitude, indicative of 
two distinct relaxation processes (see also \cite{Landsberg2009}). 

To test whether the cable exerted forces inwards, we performed a full 
ablation of the cable (Fig.~\ref{fig:ablation:global}): the entire 
edge of the wound retracts, with a single relaxation time longer than a minute.
We thus hypothesize that for local ablation
the longer time scale arises from relaxation 
at the scale of the tissue whereas the shorter one
pertains to the linear retraction of the cable. 
After ablation, the circumferential cable re-assembles on a time scale
of the order of $10$~min.

Together these observations show that the cable exerted inward forces 
and thus could contribute to force generation during closure.

\begin{figure}[!t]
\includegraphics[width=\linewidth]{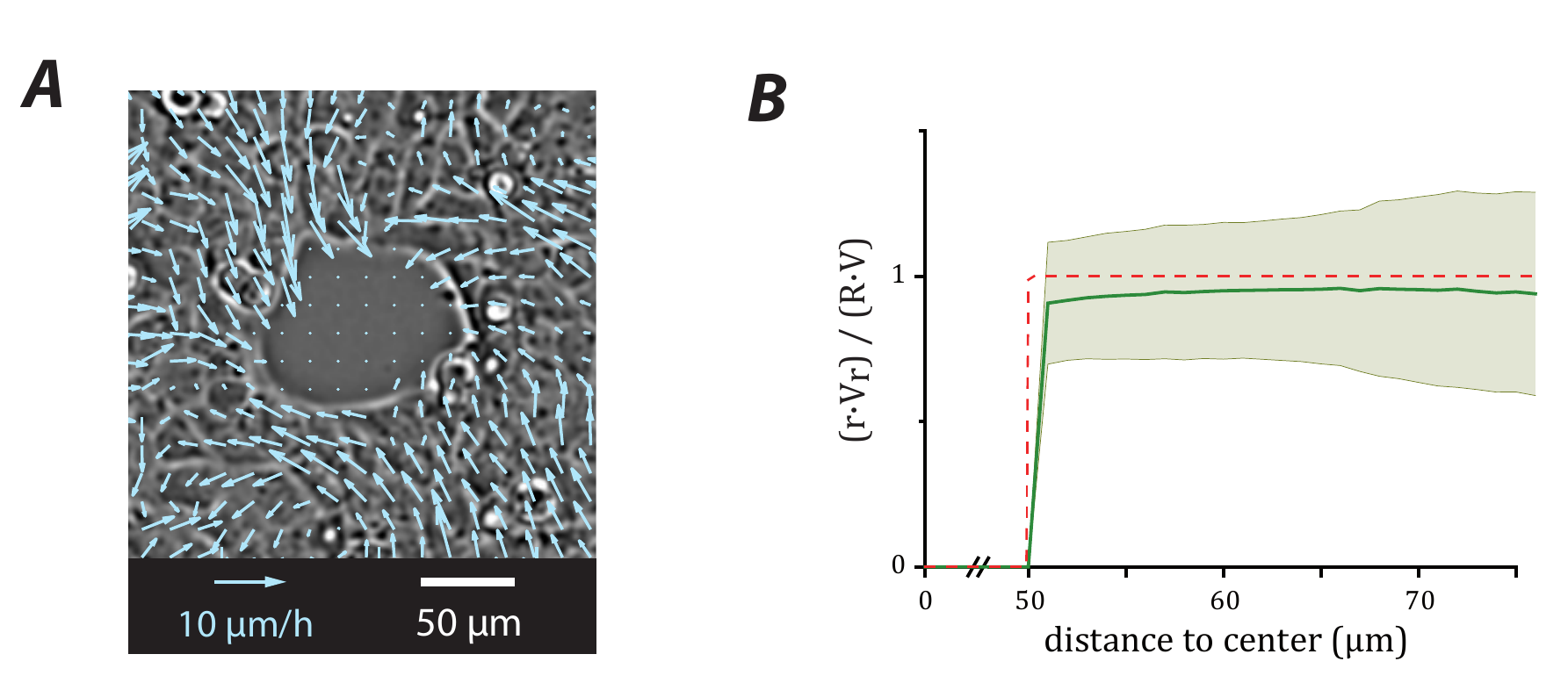}
\caption{\label{fig:tissue:velocity} 
\textbf{Epithelial velocity field}\\
\textbf{A:} Snapshot of the epithelial velocity field
(wild-type MDCK wound, $R_{\mathrm{w}} = 50 \, \mu$m, $t = 1$~h).
\\
\textbf{B:} Plot of the ensemble-averaged ratio  
$\left< \frac{r \, V_r}{R \, V} \right>_N$ vs. radius $r$
(solid green line, $N = 21$), where $V_r$ is the 
angle-averaged radial velocity component, and $R(t)$ and 
$V(t) = {\dot R}(t)$ respectively denote
the effective margin radius and velocity.
The shaded area gives the average value $\pm$ standard deviation. 
The radial velocity profile of an incompressible epithelial flow
reads  $r V_r(r,t) / (R(t)V(t))=1$ (Eq.~(\ref{eq:vr})), plotted as 
 a dashed red line for comparison.
}
\end{figure}

\subsection{Measurements at the scale of the epithelium}
\label{sec:mat:meth:tissue}

The velocity fields around the wounds were obtained through classic 
correlation-based Particle Image Velocimetry 
analysis \cite{Petitjean2010}. The center of mass of the wound was 
determined at each time with ImageJ to compute the radial averages.
Our measurement is consistent with a radial velocity component
decaying as $1/r$ (Fig.~\ref{fig:tissue:velocity}, 
see also Eq.~(\ref{eq:vr})).
 
\begin{figure}[!t]
\includegraphics[width=\linewidth]{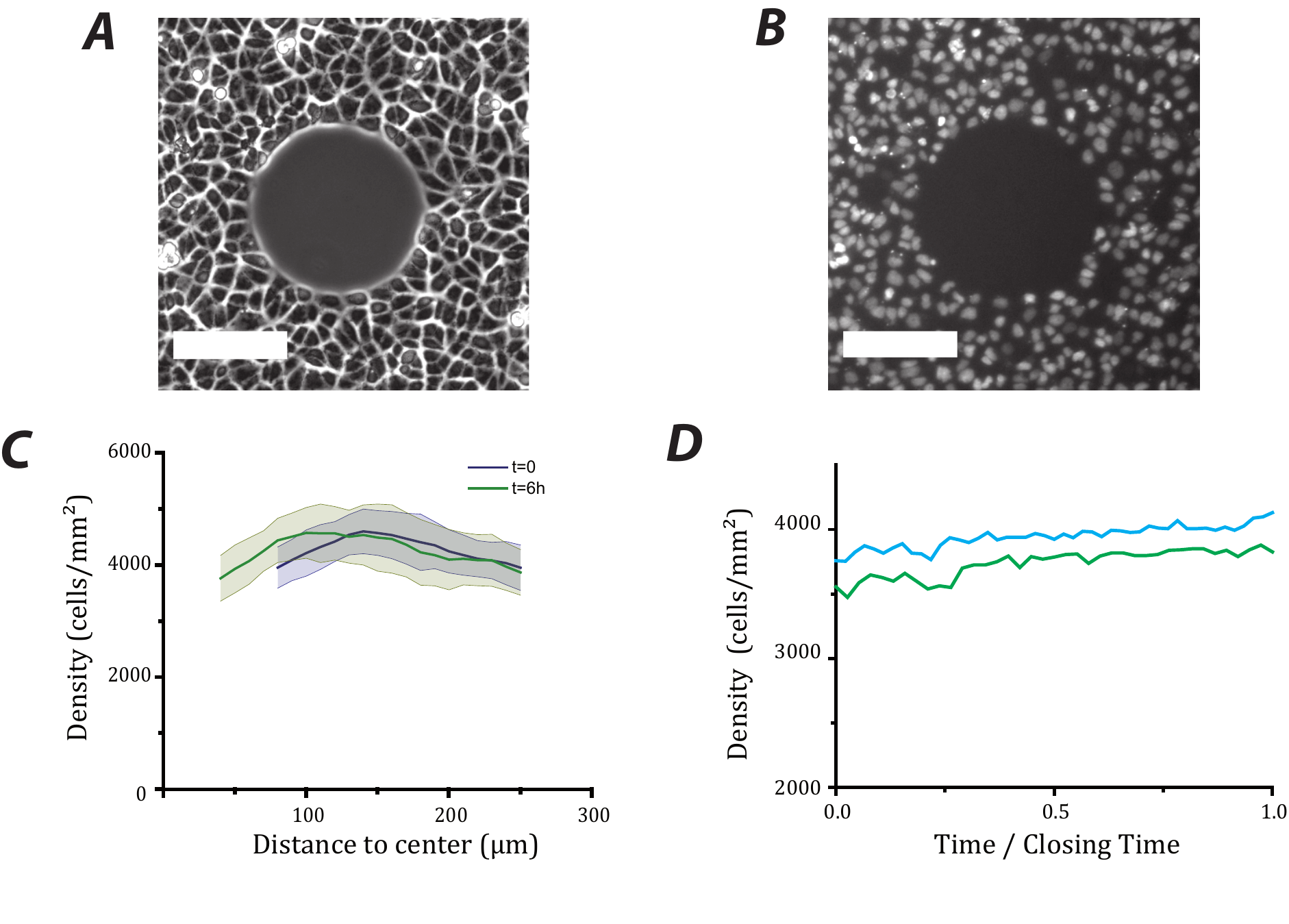}
\caption{\label{fig:tissue:density} 
\textbf{Cell number density}\\
MDCK cells expressing histone-mCherry are viewed in phase contrast (\textbf{A}) 
and in epifluorescence (\textbf{B}).
Scale bar: $100 \, \mu$m, $R_{\mathrm{w}} = 75 \, \mu$m. 
\\
\textbf{C:} Radial cell density profile at $t=0$~h and $t=6$~h.
The shaded areas give the average value $\pm$ standard deviation. 
\\
\textbf{D:} Plot of the mean cell density in the entire field of view 
as a function of time for two wounds ($R_{\mathrm{w}} = 75 \, \mu$m). 
The relative increase in density is of the order 
of $10\, \%$ during  closure.
}
\end{figure}

To measure the epithelial cell density, we used a cell line constitutively 
expressing histone-mCherry to label nuclei. Using Matlab, the position 
of the center of each nuclei was then determined. To create a density map, 
the map of the centers of the nuclei was convoluted by a 
$100 \, \mu \mathrm{m}$ x $100 \, \mu$m window while taking care 
to avoid the cell-free patches. For each pixel, we thus computed the number 
of nuclei found in this $100 \, \mu$m wide window around that pixel 
divided by the surface of the intersection of this window with the tissue.
The automated nucleus detection algorithm also allows for a simple 
count of the number of cells in the field of view at any time point.

The measured cell density is approximately uniform in space (within error
bars, Fig.~\ref{fig:tissue:density}C). It increases by less 
than $10 \%$ over the course of closure (Fig.~\ref{fig:tissue:density}D), 
consistent with typical cell division times of the order of the duration 
of closure ($\approx 10$~h) for the relevant cell 
densities \cite{Puliafito2012}.

\section{Model}
\label{sec:model}

A diversity of theoretical descriptions of wound healing have been put 
forward in the literature, with  various assumptions aimed at describing 
various conditions \cite{Nagai2009,Ouaknin2009,Bindschadler2007,Murray1988,Sherratt2002,Sadovsky2007,Almeida2009,Lee2011,Mi2007,Arciero2011}.
Cell-based descriptions include applications of the vertex \cite{Nagai2009}
and Potts \cite{Ouaknin2009} models, as well as agent-based models 
\cite{Bindschadler2007}. Early continuous descriptions include
classical biomathematical studies, reviewed in 
\cite{Murray1988,Sherratt2002}. Elastic models of a contractile
epithelium subject to  an external elastic force are investigated
in \cite{Murray1988,Sadovsky2007}, while quasistatic elasticity 
with driving border forces is treated in \cite{Almeida2009}. 
In \cite{Lee2011}, a cell monolayer is modeled as a compressible,
active (Maxwell) viscoelastic liquid with polar order.
Collective migration driven by border forces is studied
with a model of a cell layer as a compressible inviscid fluid 
including cell divisions and deaths, in one spatial dimension in \cite{Mi2007},
and in two dimensions in \cite{Arciero2011}, where the free boundary
problem is solved using a level set method.

Here, we formulate a simple continuum mechanics description of wound closure, 
where we take advantage of our experimental observations that 
(i) cell division and death are negligible during the time of wound closure, 
(ii) there is no apparent orientational order of the cells,
(iii) the wound shape remains approximately circular over the course of
the experiment, and 
(iv) the flow is incompressible.
We first detail our theoretical description 
(section~\ref{sec:model:ingredients}),
before we study three different epithelial rheologies, based on 
constitutive equations for either a simple inviscid or viscous
liquid (sections~\ref{sec:model:inviscid} and \ref{sec:model:viscous},
respectively) or an elastic solid (section~\ref{sec:model:elastic}).
Each rheology allows to obtain an analytical expression for the closure 
dynamics of the circular model wounds created by the 
experimental protocol.

\subsection{Continuum mechanics epithelization}  
\label{sec:model:ingredients}

In order to understand wound closure dynamics on the scale of the 
epithelium, we aim at describing stresses and strains 
on large length scales, as compared to the cell size. Using continuum 
mechanics, we formulate an effective two-dimensional description of 
epithelization that takes into account the macroscopic tissue material 
properties.

Conservation of cell number in the epithelium is expressed by
\begin{equation}
\del_t n + \del_\alpha (nv_\alpha) = n(k_{\rm d} - k_{\rm a}) \, ,
\end{equation}
where $n$ is the cell number density, $v_\alpha$ the tissue velocity field, 
and $k_{\rm d}$ and $k_{\rm a}$ are the rates of cell division and cell death, 
respectively. By convention, greek indices denote vector components, and 
are summed when repeated. 
We assume $k_{\rm d}=k_{\rm a}=0$ in the following, consistent with our 
experimental observations that both cell division and cell death 
are negligible during the time course of wound closure. 
Furthermore, the cell number density $n=n_0$ is approximately constant
(Fig.~\ref{fig:tissue:density}).
The cell number balance equation then becomes a constraint on the tissue 
flow field, $\del_\alpha v_\alpha=0$: the flow is incompressible 
(Fig.~\ref{fig:tissue:velocity}). 

In a continuous material, mechanical forces are balanced locally if inertial 
terms can be neglected, as is the case here. 
Force balance is then expressed as 
\begin{equation}
\label{eq:force_balance}
\del_\beta\sigma_{\ab} = - f^{\rm ext}_\alpha \, ,
\end{equation}
where internal forces are described by 
the stress tensor $\sigma_{\ab}$, and $f^{\rm ext}_\alpha$ 
denotes external forces. Here, the external force is due to friction 
with the substrate, and with $\xi$ being a friction coefficient we write 
$f^{\rm ext}_\alpha = - \xi v_\alpha$.
Together with a constitutive equation for the stress tensor and appropriate 
boundary conditions, Eq.~\eqref{eq:force_balance} allows to solve for the 
deformation and cell flow field in the epithelium. 
The constitutive equation for the stress tensor 
accounts for the tissue material properties. In general, the stress tensor can 
be decomposed into an isotropic part $\sigma$ and a deviatoric (traceless) 
part $\tilde\sigma_{\ab}$ according to
\begin{equation}
\label{eq:stress_decomp}
\sigma_{\ab} = \sigma \delta_{\ab} + \tilde \sigma_{\ab} \, ,
\end{equation}
where $\delta_{\ab}$ denotes Kronecker's symbol, 
and $\tilde\sigma_{\alpha\alpha}=0$ by definition.

In the following, we consider an epithelium where a model wound 
with initial radius $R_0$ is created at $t_0=0$, centered about the origin $O$.
We assume that the circular shape is preserved during the closure process 
and denote by $R(t)$ the wound radius at time~$t$ (Fig.~3A).
The wound closes because of forces exerted at the margin, 
either by actively pulling cells or by an acto-myosin cable 
that spans over the whole perimeter. 
Using polar coordinates, the stress boundary condition at the margin reads
\begin{equation}
\label{eq:radial_stress_bc}
\sigma_{rr}|_{R(t)} =   \sigma_{\rm p} + \frac{\gamma}{R}\, ,
\end{equation}
where $\sigma_{\rm p}$ is a protrusive stress that 
accounts for forces exerted by the cells at the wound margin,
and $\gamma$ is a tension that describes purse-string forces due to an 
acto-myosin cable around the wound. 
Introducing the length scale $R_{\gamma}=\gamma/\sigma_{\rm p}$,
we expect that the purse-string mechanism (resp.~the protrusive forces)
will dominate the dynamics at scales smaller (resp.~larger) than $R_{\gamma}$. 

Assuming rotational invariance of the flow allows to express 
the velocity field as $\vec{v} = v_r(r,t)\:\vec{e}_r \,$, where the 
non-vanishing radial component depends only on the distance $r$ relative 
to the center $O$ of the initial wound. Using the incompressibility 
constraint $\nabla\cdot\vec{v}=0$, we obtain
$v_r(r,t) = A(t)/r$,
where $A(t)$ can be determined from the 
kinematic boundary condition at the margin. Since 
$v_r(r = R(t),t)=\dot R(t)$,
we can express $v_r(r,t)$ in terms of $r$ and the wound radius $R(t)$ only
\begin{equation}
  \label{eq:vr}
v_r(r,t) = \frac{R(t) \dot R(t)}{r} \, .  
\end{equation}
Using this expression with Eqs.~\eqref{eq:force_balance} and 
\eqref{eq:radial_stress_bc} allows 
to find a dynamical equation for the wound radius $R(t)$. 
In the following sections, we derive and solve this dynamical equation---or 
rather the inverse problem t = t(R)---for three different constitutive 
equations, each highlighting a different epithelial rheology.

\subsection{Inviscid fluid}
\label{sec:model:inviscid}

For simplicity, we first assume that the epithelium behaves as an 
incompressible, inviscid fluid on the relevant time and length scales. 
In this case, the stresses are purely isotropic and do not depend on 
tissue viscosity or elasticity.  In the incompressible limit, the 
isotropic part of the stress becomes a Lagrange multiplier which is 
determined from the mechanical boundary conditions, and we simply 
write $\sigma=-P$. The stress tensor thus reads
\begin{equation}
  \label{eq:stress_inviscid}
  \sigma_{\alpha\beta} = - P \, \delta_{\alpha\beta} \,,
\end{equation}
where $P$ is the pressure field at the scale of the epithelium. 
Using rotational invariance ($P = P(r,t)$) and Eq.~\eqref{eq:vr} 
for the velocity field, the force balance~\eqref{eq:force_balance} becomes 
$ \del_r P  =  -\xi R \dot R/r \,$.
The pressure follows as $P = - \xi R \dot R \ln r + C$, where $C = C(t)$ is 
a function of time.
Note that in principle, $C(t)$ is determined by the boundary condition at 
$r\to\infty$, which is an ill-defined limit in two dimensions. 
We therefore introduce a constant, long-range cut-off $R_{\max}$ 
at which the pressure vanishes and write
\begin{equation}
\label{eq:incompr_pressure_invisc}
P(r,t) = - \xi R(t) \dot R(t) \ln  \frac{r}{R_{\max}}  \, .
\end{equation}
Since $\dot R(t) \le 0$ and $r \le R_{\max}$, the pressure is negative:
the epithelium is under tension.

A dynamical equation for the wound radius $R(t)$ follows from the 
stress boundary condition at the margin, Eq.~\eqref{eq:radial_stress_bc}, and with the above expression for $P$ we find
\begin{equation}
\label{eq:R_dynamics_invisc}
\xi R \ln{\left( \frac{R}{R_{\max}} \right)} \, \dot R  = \sigma_{\rm p} +  
\frac{\gamma}{R} \, .
\end{equation}
Using the characteristic length $R_{\gamma}=\gamma/\sigma_{\rm p}$,
we rewrite the evolution equation as
\begin{equation}
\label{eq:valp}
{\rm d}t = \frac{\xi}{\sigma_{\rm p}} \, 
\frac{R^2 }{R + R_{\gamma}} \ln{\left(\frac{R}{R_{\max}}\right)} \, {\rm d}R \, .
\end{equation}
Integration yields the function $t(R)=\tilde t(R) - \tilde t(R_0)$,
with
\begin{multline}
\label{eq:ttilde_invisc}
4 D \, \tilde t(R) = 
   - R^2 \, \left( 1 + 2 \ln \frac{R_{\mathrm{max}}}{R} \right) \\
 + \, 4 R_{\gamma} R \left( 1 + \ln{\frac{R_{\max}}{R}} \right) 
+ \, 4 R_{\gamma}^2
\left(  \text{Li}_2(-\frac{R}{R_{\gamma}}) -
\ln{\frac{R_{\max}}{R}} \ln{\frac{R+R_{\gamma}}{R_{\gamma}}} 
    \right)   \, .
\end{multline}
Here, we introduce the epithelization coefficient
$D = \sigma_{\rm p}/\xi$, which has the dimension of a diffusion coefficient, and Li$_2$ stands for the dilogarithm function defined as
${\rm Li}_2(x) = \sum_{k=1}^\infty x^k/k^2$. 
The integration constant is determined by the initial condition $t(R_0)=0$. 
Since $\text{Li}_2(0) = 0$, the closure time is finite:
$t_{\rm c} \equiv t(R = 0) = \tilde t(0) - \tilde t(R_0) = - \tilde t(R_0)$.

When the contribution of the acto-myosin cable is negligible, $R_{\gamma}\to0$,
the expression for $t(R)$ simplifies to
\begin{equation}
  \label{eq:inviscid:t:R}
  t(R) \simeq \frac{R_0^2}{4 D} \, \left( 1 + 
2 \ln\frac{R_{\mathrm{max}}}{R_0}
\right)
- \frac{R^2}{4 D} \, \left( 1 + 
2 \ln \frac{R_{\mathrm{max}}}{R}
\right),
\end{equation}
and the closure time follows as
\begin{equation}
  \label{eq:inviscid:tc:R0}
  t_{\rm c}(R_0) \simeq \frac{R_0^2}{4 D} \, \left( 1 + 
2 \ln \frac{R_{\mathrm{max}}}{R_0}  \right)
\end{equation}
in the same limit.
This result implies that under the above assumptions, \emph{i.e.}, 
for an inviscid epithelium, the closure of a circular model wound completes
in a finite time, independently of whether a contractile cable contributes 
to force production or not.

\subsection{Viscous fluid}
\label{sec:model:viscous}

Taking into account viscous stresses, the deviatoric stress tensor
is given by
\begin{equation}
\tilde\sigma_{\alpha\beta} = 2\eta \tilde v_{\alpha\beta} \, ,
\end{equation}
where $\eta$ is an effective tissue shear viscosity and 
$\tilde v_{\alpha\beta}$ is the traceless part of the velocity gradient tensor 
$v_{\alpha\beta}=\tfrac{1}{2}(\del_\alpha v_\beta + \del_\beta v_\alpha)$. 
The isotropic part of the stress becomes again a Lagrange multiplier, 
and we write $\sigma=-P$ as before. Incompressibility also implies 
that $v_{\gamma\gamma}=0$, and thus $\tilde v_{\alpha\beta} = v_{\alpha\beta}$.

Taking into account rotational invariance,
the radial component of the force balance~\eqref{eq:force_balance} reads
\begin{equation}
\label{eq:force_balance_rot_symm}
\del_r \sigma + \del_r \tilde\sigma_{rr} + 2 \frac{\tilde\sigma_{rr}}{r} 
= \xi v_r \, .
\end{equation}
Inserting the constitutive equations, we obtain as before 
$\del_r P = -\xi R \dot R/r$. 
Expression~\eqref{eq:incompr_pressure_invisc}
for the pressure field is therefore unchanged.
With $\sigma_{rr}=-P + 2\eta \, \del_r v_r$, the boundary 
condition~\eqref{eq:radial_stress_bc} now leads to
\begin{equation}
\label{eq:R_dynamics_visc}
\dot R  = \frac{\gamma + \sigma_{\rm p} R}{ \xi R^2 \ln{R/R_{\max}} - 2\eta}   \, .
\end{equation}
Introducing the length scale $R_{\eta} = \sqrt{\eta/\xi}$, integration yields
\begin{equation}
  \label{eq:viscous:t:R:1}
t(R) = \tilde t(R) - \tilde t(R_0) \, ,
\end{equation}
with
\begin{multline}
  \label{eq:viscous:t:R:2}
4 D \, \tilde t(R) = 
- R^2 \left( 1 + 2  \ln{\frac{R_{\max}}{R}} \right)  
+ 8 R_{\eta}^2 \, \ln{\frac{R_{\max}}{R + R_{\gamma}}} \\
+  4 R_{\gamma} R \, \left( 1 + \ln{\frac{R_{\max}}{R}} \right) 
+  4 R_{\gamma}^2
\left( 
\text{Li}_2(-\frac{R}{R_{\gamma}})   -
\ln{\frac{R_{\max}}{R}} \ln{\frac{R+R_{\gamma}}{R_{\gamma}}} 
\right)   \, .
\end{multline}
In the limit of vanishing viscosity, $R_{\eta} \to 0$, the above expression reduces to Eq.~\eqref{eq:ttilde_invisc}, consistent with the assumption of vanishing deviatoric stresses in the inviscid case.

The closure time is again finite,
 $t_{\rm c} \equiv t(R = 0) = \tilde t(0) - \tilde t(R_0) = 
2 (\eta/\sigma_{\rm p}) \, \ln(R_{\max}/R_{\gamma}) - \tilde t(R_0)$,
and tends to  expression~\eqref{eq:inviscid:tc:R0} in the limit
where both $R_{\eta}$ and $R_{\gamma}$ are negligible.
However, if $R_{\eta}$ remains finite, the closure time diverges 
in the limit $R_{\gamma} \to 0$. The model predicts that,
in the absence of a contractile cable, 
circular model wounds do not complete closure in finite time when 
viscous stresses in the epithelium cannot be neglected.
This somewhat surprising result is an artifact of the continuous description:
in fact closure will complete, thanks to cell-scale mechanisms not
taken into account by the model, as soon the wound radius is smaller than a 
microscopic cut-off length $a$, with a finite closure time of the 
order of $\tilde t(a) - \tilde t(R_0)$.

\subsection{Elastic solid}
\label{sec:model:elastic}

When deformations are small,
the constitutive equation for an incompressible elastic material reads
\begin{equation}
\tilde\sigma_{\alpha\beta} = 2\mu \tilde u_{\alpha\beta} \, ,
\end{equation}
where $\mu$ is the shear elastic modulus and $\tilde u_{\alpha\beta}$ 
is the traceless part of the strain tensor. The latter is defined as 
$u_{\alpha\beta}=\tfrac{1}{2}(\del_\alpha u_\beta + \del_\beta u_\alpha)$ 
for a displacement field $u_\alpha$.
Incompressibility implies that $u_{\gamma\gamma}=0$, and thus 
$\tilde u_{\alpha\beta} = u_{\alpha\beta}$. In this limit, 
the isotropic stress becomes again a Lagrange multiplier and 
we write $\sigma=-P$.

In the case of rotational invariance, we can express the elastic displacement
field as $\vec{u}=u_r(r,t) \, \vec{e}_r$. 
Using the incompressibility condition $\nabla\cdot\vec{u}=0$
together with the boundary condition $u_r(R,t) = R(t)-R_0$, 
we obtain $u_r$ as a function of $r$ and $R(t)$, 
\begin{equation}
u_r(r,t) = \frac{R(t)(R(t)-R_0)}{r} \, .
\end{equation}
One can check that this expression verifies
$\dot R(t) \equiv \left( \del_t  + v_r \del_r \right) u_r(r = R(t), t) = v_r(r = R(t), t)$
at all times $t\ge0$.
The differential equation for $P$ resulting from force balance
is again unchanged,  $P$ is given by Eq.~\eqref{eq:incompr_pressure_invisc}. 
Since the radial stress in the epithelium 
is given by
\begin{equation}
  \label{eq:elastic:radialstress}
  \sigma_{rr}  = - P + 2 \mu \del_r u_r \, ,
\end{equation}
the stress boundary condition~\eqref{eq:radial_stress_bc}
yields the following dynamical equation for the wound radius $R(t)$
\begin{equation}
\label{eq:R_dynamics_incompr_elastic}
\dot R = \frac{\gamma + \sigma_{\rm p} R + 2\mu(R-R_0)}{\xi R^2 \ln{R/R_{\max}}}  \, .
\end{equation}
Formally, elastic restoring forces and forces driving 
epithelization balance at the equilibrium radius $R_{\rm e}$ with
\begin{equation}
\label{def:Re}
R_{\rm e} = \frac{2\mu R_0 - \gamma}{\sigma_{\rm p}+2\mu} \, .
\end{equation}
Taking into account the initial condition $t(R_0)=0$, integration of
Eq.~\eqref{eq:R_dynamics_incompr_elastic} yields
\begin{equation}
\label{eq:solution_incompressible_elastic_0}
t(R) = \tilde t(R) - \tilde t(R_0) \, ,
\end{equation}
where $\tilde t(R)$ is given by
\begin{multline}
\label{eq:solution_incompressible_elastic}
4 D_S \, \tilde t(R) = 
- R^2  - 4 R R_{\mathrm e} + 2 R \left( R + 2 R_{\mathrm e} \right)  \ln{\frac{R}{R_{\max}}} 
\\
+ \, 4 R_{\mathrm e}^2 \left( \ln{\frac{R}{R_{\max}}} \ln{(1-\frac{R}{R_{\mathrm e}})} +  \text{Li}_2(\frac{R}{R_{\mathrm e}})   \right) \, . 
\end{multline}
Here, $D_{\rm S} = \frac{ \sigma_{\rm p} + 2\mu}{\xi}$ has the dimension 
of a diffusion coefficient, and differs from the epithelization
coefficient $D$ by a factor of $(1 + \frac{2 \mu}{\sigma_{\rm p}})$. 
In the limit of vanishing elastic modulus 
($2 \mu \ll \sigma_{\rm p}$ and  $2 \mu \ll \gamma/R_0$), 
expression~\eqref{eq:solution_incompressible_elastic} for $t(R)$ reduces to
the one obtained for an inviscid fluid as given by Eq.~\eqref{eq:ttilde_invisc}.
Of course only positive values of the radius 
are physical and closure stops when $R(t_c) = 0$. 

The above result for $t(R)$ implies that the wound closure eventually 
completes whenever $R_{\rm e} \le 0$. This is the case for large enough values 
of the line tension $\gamma$, \emph{i.e.}, $\gamma \ge 2\mu R_0$. 
The closure time is then given by 
$t_{\rm c}= t(R = 0) = \tilde t(0) - \tilde t(R_0) = - \tilde t(R_0)$.
In the particular case where line tension and elasticity balance
exactly, $\gamma = 2\mu R_0$ and thus $R_{\rm e} = 0$, 
Eq.~\eqref{eq:solution_incompressible_elastic} reduces to 
Eq.~\eqref{eq:inviscid:t:R}, and the closure time follows
as given by Eq.~\eqref{eq:inviscid:tc:R0} with the substitution $D \to D_S$.

When the equilibrium radius is positive but small, 
$0 < R_{\mathrm e} \simeq a$,  where $a$ is of the order of the size of 
a cell, epithelization may proceed to a scale small enough 
that microscopic mechanisms, not accounted for within 
the continuous description, terminate the epithelization process. 
This might be the case even for small values of the line tension
$0 \le \gamma \le 2\mu R_0$ provided that the protrusive stress 
dominates the elastic modulus, $\sigma_{\rm p}\gg 2\mu$
(see Eq.~\eqref{def:Re}).

When the epithelial elastic modulus is large enough
($2 \mu \sim \sigma_{\rm p}$ and  $2 \mu > \gamma/R_0$), the equilibrium radius 
is strictly positive: wound closure halts due to elastic forces.
Expression~\eqref{eq:solution_incompressible_elastic} 
takes complex values when $R_{\rm e}>0$. However, the identity
\begin{equation}
\text{Li}_2(x) + \text{Li}_2(1-x) + \ln{(1-x)}\ln{x} = \frac{\pi^2}{6} \,
\end{equation}
allows to rewrite $t(R)$ as
\begin{multline}
\label{solution_incompressible_elastic_positive}
t(R) = \frac{\xi}{4 \left( \sigma_{\rm p} + 2\mu \right)}  \left[  (R_0^2- R^2)  + 4 (R_0-R) R_{\mathrm e} + 2 R \left( R + 2 R_{\mathrm e} \right)  \ln{\frac{R}{R_{\max}}} \right. \\
\left.  - \, 2 R_0 \left( R_0 + 2 R_{\mathrm e} \right)  \ln{\frac{R_0}{R_{\max}}} +  4 R_{\mathrm e}^2 \left(  \text{Li}_2(1-\frac{R_0}{R_{\mathrm e}}) - \text{Li}_2(1-\frac{R}{R_{\mathrm e}}) + \ln{\frac{R-R_{\mathrm e}}{R_0-R_{\mathrm e}}}  \ln{\frac{R_{\mathrm e}}{R_{\max}}} \right)  \right] \, ,
\end{multline}
where all terms are real-valued for $R>R_{\mathrm e}>0$.
In this case the closure time is infinite.

\begin{figure}[!h]
\includegraphics[width=0.8\linewidth]{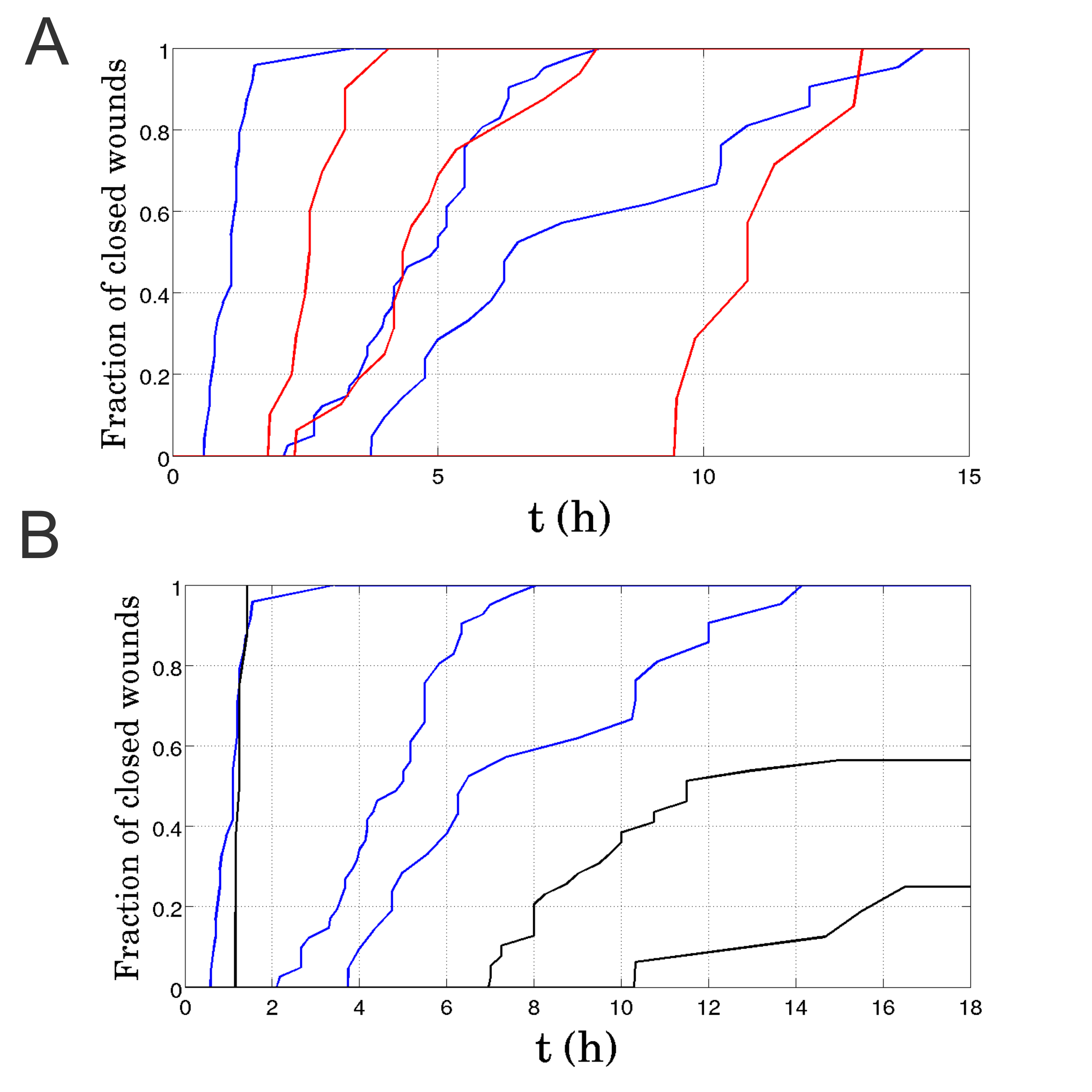}
\caption{\label{fig:MDCK:inhib} 
\textbf{Cumulative distributions of closure times for MDCK wounds.}\\
\textbf{A:} comparison between wild-type (blue curves, $N=24$, $41$ and 
$16$ respectively) and Rho$^-$ assays (red curves, $N=10$, $16$ and 
$7$ respectively), $R_{\mathrm{w}} = 25 \, \mu$m, $50 \, \mu$m and $100 \, \mu$m
from left to right.
\\ 
\textbf{B:}
comparison between wild-type (blue curves, same data as in A) and
Rac$^-$ assays (red curves, $N=8$, $39$ and $16$ respectively), 
same sizes from left to right.
A fraction of the Rac$^-$ wounds do not complete closure within
the observation time $t = 18$~h.  
}
\end{figure}

\section{Data analysis}
\label{sec:data}

\begin{figure}[!t]
\textbf{A}
\includegraphics[width=7.7cm]{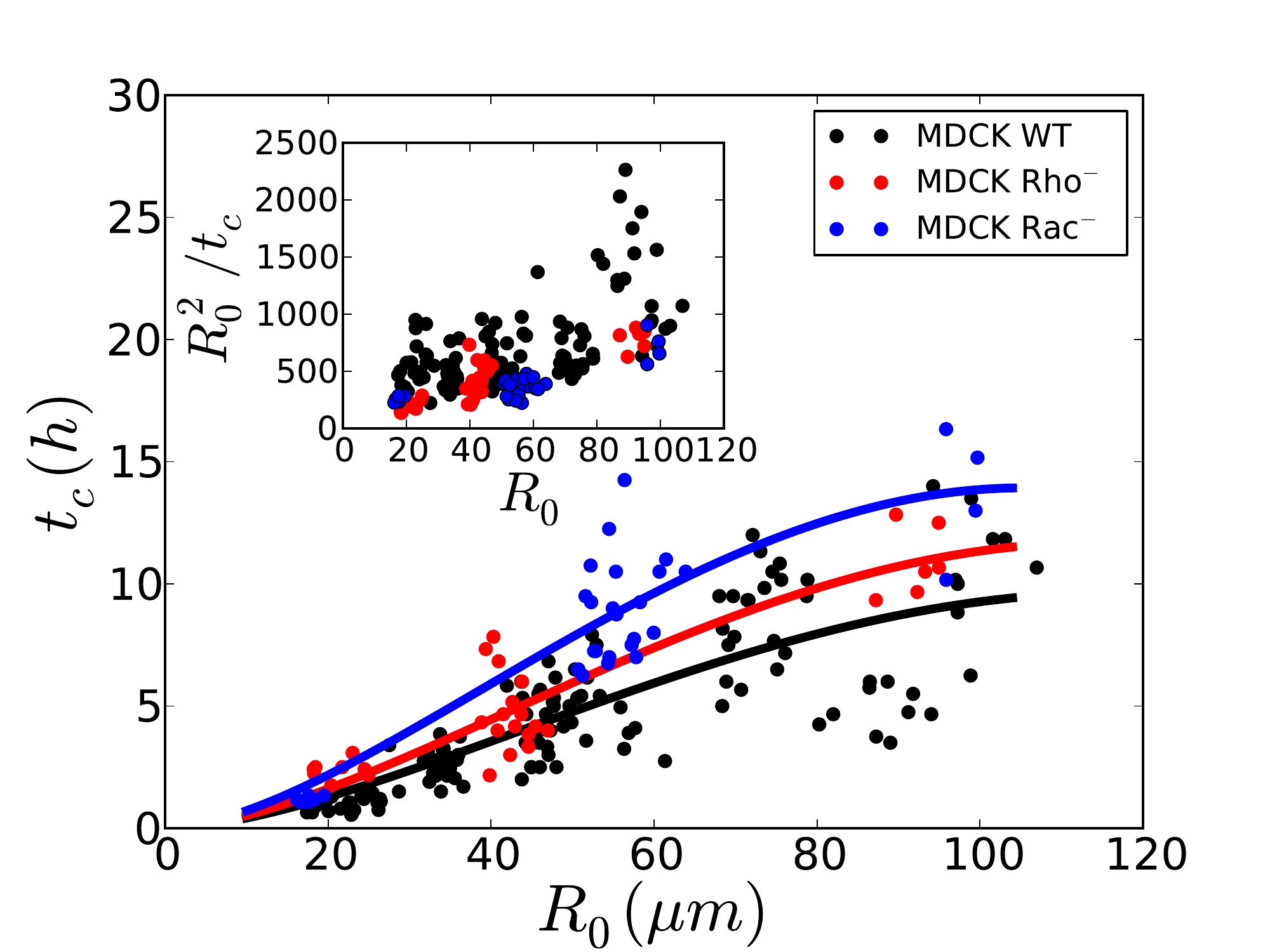}
\textbf{B}
\includegraphics[width=7.7cm]{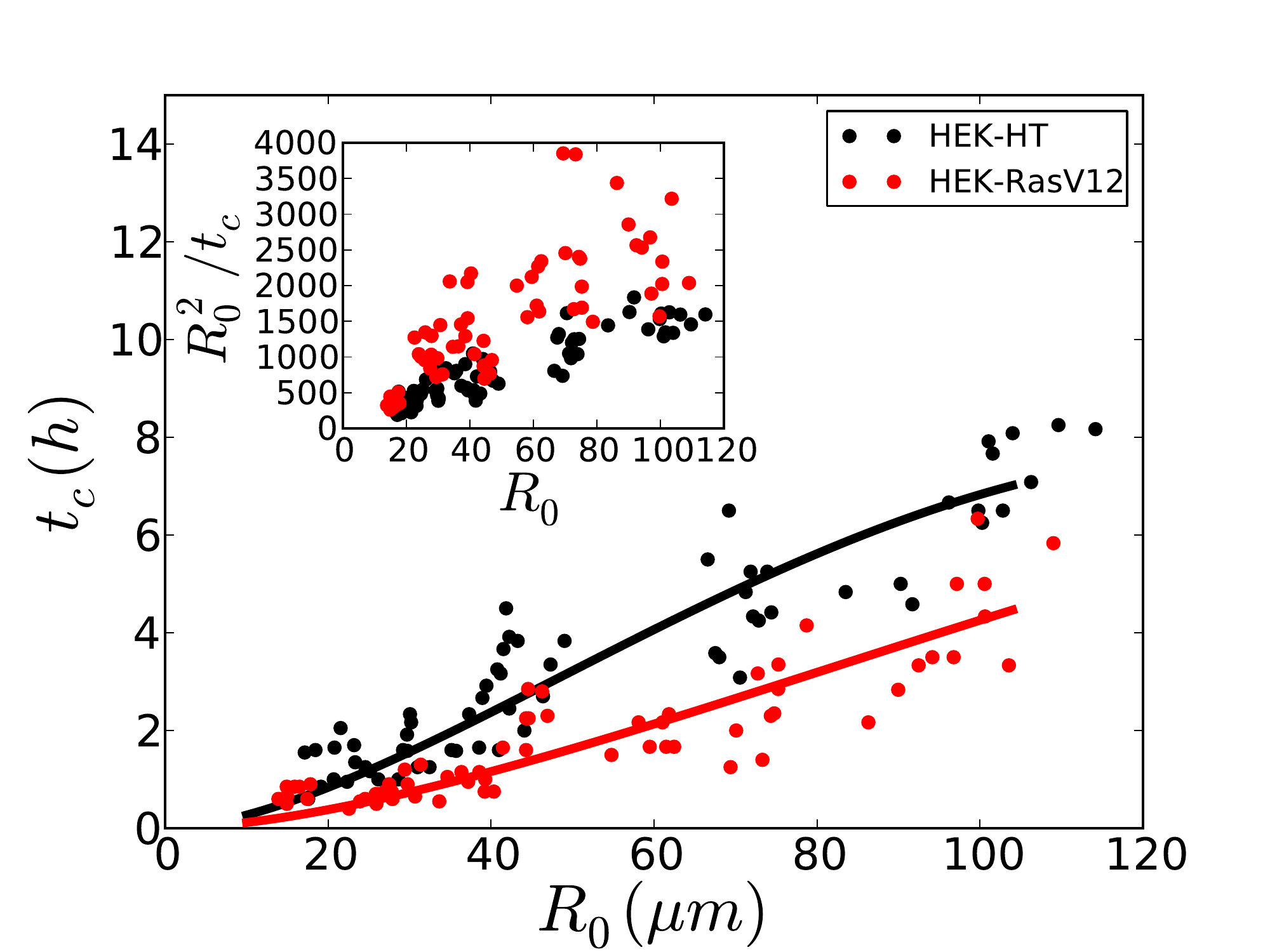}
\caption{\label{fig:tc:inviscid} 
\textbf{Inviscid fluid:} Closure time $t_{\mathrm c}$ as a function 
of the initial effective radius $R_0$ (circles), 
fitted by Equation~(\ref{eq:tc:R0:inviscid:nocable}) (solid curves)
with the constraints $D, R_{\mathrm{max}} \ge 0$. 
One dot corresponds to one wound.
\\ 
\textbf{A:} MDCK wounds. Wild Type, Rho$^-$ and  Rac$^-$ assays.
The physical parameters of epithelization $(D, R_{\mathrm{max}})$
are given within a $95 \, \%$ confidence interval.
MDCK WT ($N = 130$):
$\sigma_{\rm p}/\xi = 353 \pm 38 \, \mu \mathrm{m}^2 \, \mathrm{h}^{-1}$,
$R_{\mathrm{max}} = 117 \pm 11 \, \mu \mathrm{m}$; 
MDCK Rho$^-$ ($N = 30$):
$\sigma_{\rm p}/\xi = 278 \pm 40 \, \mu \mathrm{m}^2 \, \mathrm{h}^{-1}$,
$R_{\mathrm{max}} = 114 \pm 14 \, \mu \mathrm{m}$; 
MDCK Rac$^-$ ($N = 34$):
$\sigma_{\rm p}/\xi = 198 \pm 22 \, \mu \mathrm{m}^2 \, \mathrm{h}^{-1}$,
$R_{\mathrm{max}} = 105 \pm 9 \, \mu \mathrm{m}$; 
\\
\textbf{B:} HEK-HT and HEK-RasV12 wounds. 
HEK-HT ($N = 63$):
$\sigma_{\rm p}/\xi = 572 \pm 57 \, \mu \mathrm{m}^2 \, \mathrm{h}^{-1}$,
$R_{\mathrm{max}} = 132 \pm 12 \, \mu \mathrm{m}$; 
HEK-RasV12 ($N = 65$):
$\sigma_{\rm p}/\xi = 1531 \pm 363 \, \mu \mathrm{m}^2 \, \mathrm{h}^{-1}$,
$R_{\mathrm{max}} = 223 \pm 77 \, \mu \mathrm{m}$.\\
\textbf{Insets}: for all cell types and conditions, 
the ratio of initial effective area over
closure time $R_0^2/t_{\mathrm c}$ increases with initial radius $R_0$.
}
\end{figure}

In section~\ref{sec:model}, we obtained analytical expressions of 
$t_{\mathrm c}(R_0)$ and $t(R)$, corresponding to different epithelial
rheologies. In order to estimate the physical parameters of the epithelia, 
we now fit experimental data by these expressions, using the 
Levenberg-Marquardt algorithm for nonlinear
least-squares fitting, implemented in \textsc{Python} \cite{Note2}.
We successively examine fits of closure times vs.~initial effective radii 
in section~\ref{sec:data:tc} and  fits of individual trajectories $R(t)$
in section~\ref{sec:data:traj}.
Finally, we discuss the values of physical parameters
thus estimated in section~\ref{sec:data:param}.

\subsection{Closure time data}
\label{sec:data:tc}

The closure time is a robust quantity that depends only weakly
on the image analysis method: at a given time $t$, the wound is either 
open or closed. The experimental uncertainty on $t_{\mathrm c}$ is of the order
of the time resolution of data acquisition, between $3$ and $15$ minutes
depending on the size of the wound. Fig.~\ref{fig:MDCK:inhib} gives the
empirical cumlative distribution functions of closure times
for MDCK wounds, including the effect of inhibitors.

Fig.~\ref{fig:tc:inviscid} shows that closure time data 
pertaining to all cell types and conditions is well fitted by 
Equation~(\ref{eq:inviscid:tc:R0}),
obtained for an inviscid epithelium without cable. 
Of note, experimental data plateaus for $R_0 \gtrsim 100 \, \mu$m.
This behavior is not consistent with a simple scaling 
relationship where the closure time would be proportional
to the initial area $\pi R_0^2$, as proposed in \cite{Anon2012}.

We now ask whether this simple description is robust, and consider
this question in the case of MDCK wild-type wounds, 
for which the number of wounds is largest ($N = 130$). 
As shown in section~\ref{sec:model}, different assumptions made on the
epithelial rheology lead to different expressions of the closure 
time $t_{\mathrm c}$ as a function of the initial radius $R_0$. 
Although an inviscid epithelium may close without cable, 
strictly speaking, both a viscous and an elastic epithelium 
require a finite line tension
($\gamma \neq 0$) for closure to reach completion.

For convenience, we summarize below the analytical expressions 
obtained for $t_{\mathrm c}(R_0)$:
\begin{itemize}
\item[-]{\emph{inviscid liquid, without cable} ($\gamma = 0$, 
$D = \sigma_{\rm p}/\xi$):
\begin{equation}
  \label{eq:tc:R0:inviscid:nocable}
4 D \,  t_{\mathrm c}(R_0) = R_0^2 \, \left( 1 + 
2 \ln \frac{R_{\mathrm{max}}}{R_0}  \right)
\end{equation}
} 
\item[-]{\emph{inviscid liquid, with cable} ($\gamma \neq 0$, 
$R_{\gamma}=\gamma/\sigma_{\rm p}$):
\begin{multline}
  \label{eq:tc:R0:inviscid:cable}
4 D \, t_{\mathrm c}(R_0) = 
R_0^2 \left( 1 + 2  \ln{\frac{R_{\max}}{R_0}} \right)  
-  4 R_0 R_{\gamma} \, \left( 1 + \ln{\frac{R_{\max}}{R_0}} \right) \\
-  4 R_{\gamma}^2
\left( 
\text{Li}_2(-\frac{R_0}{R_{\gamma}})   +
\ln{\frac{R_{\max}}{R_0}} \ln{\frac{R_{\gamma}}{R_0+R_{\gamma}}} 
\right)   
\end{multline}
} 
\item[-]{\emph{viscous liquid, with cable} ($R_{\eta} = \sqrt{\eta/\xi}$):
\begin{multline}
  \label{eq:tc:R0:viscous:cable}
4 D \, t_{\mathrm c}(R_0) = 
R_0^2 \left( 1 + 2  \ln{\frac{R_{\max}}{R_0}} \right)  
-  4 R_0 R_{\gamma} \, \left( 1 + \ln{\frac{R_{\max}}{R_0}} \right) \\
+ 8 R_{\eta}^2 \, \ln{\frac{R + R_{\gamma}}{R_{\gamma}}} 
-  4 R_{\gamma}^2
\left( 
\text{Li}_2(-\frac{R_0}{R_{\gamma}})   +
\ln{\frac{R_{\max}}{R_0}} \ln{\frac{R_{\gamma}}{R_0+R_{\gamma}}} 
\right)   
\end{multline}
} 
\item[-]{\emph{elastic solid, with cable} 
($R_{\mathrm e} = \frac{2\mu R_0 - \gamma}{\sigma_{\rm p}+2\mu} \le 0$,
$D_S = \frac{ \sigma_{\rm p} + 2\mu}{\xi}$):
\begin{multline}
  \label{eq:tc:R0:elastic:cable}
4 D_S \, t_{\mathrm c}(R_0) =   R_0^2 \left( 1 + 2  \ln{\frac{R_{\max}}{R_0}} \right)   
+ 4 R_0 R_{\mathrm e} \, \left( 1 + \ln{\frac{R_{\max}}{R_0}} \right) \\
 - \, 4 R_{\mathrm e}^2 \left( \text{Li}_2(\frac{R_0}{R_{\mathrm e}})
- \ln{\frac{R_{\max}}{R_0}} \ln{(1-\frac{R_0}{R_{\mathrm e}})} 
   \right) 
\end{multline}
} 
\end{itemize}

\begin{figure}[!t]
\textbf{A}
\includegraphics[width=7.7cm]{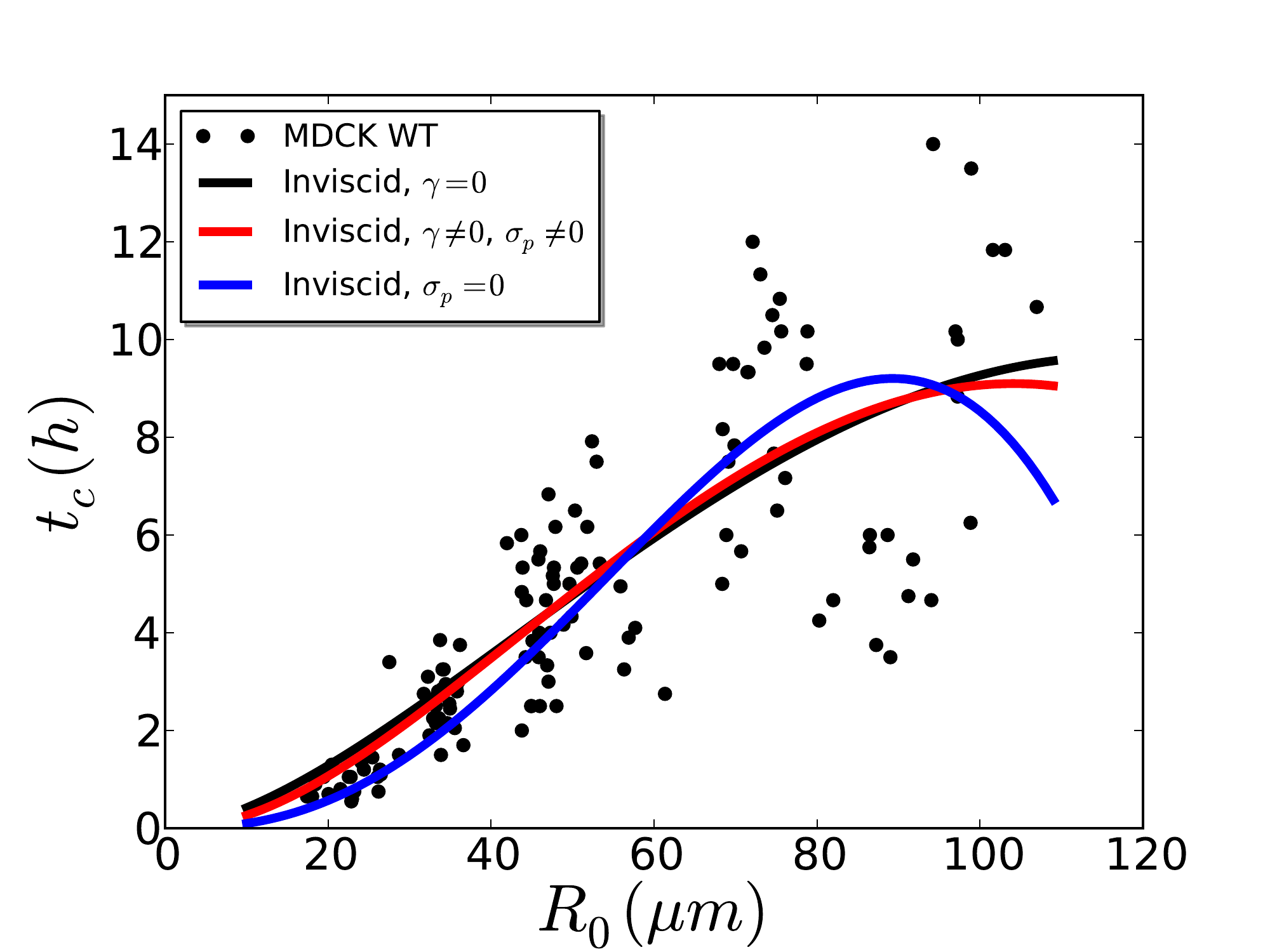}
 \textbf{B}
\includegraphics[width=7.7cm]{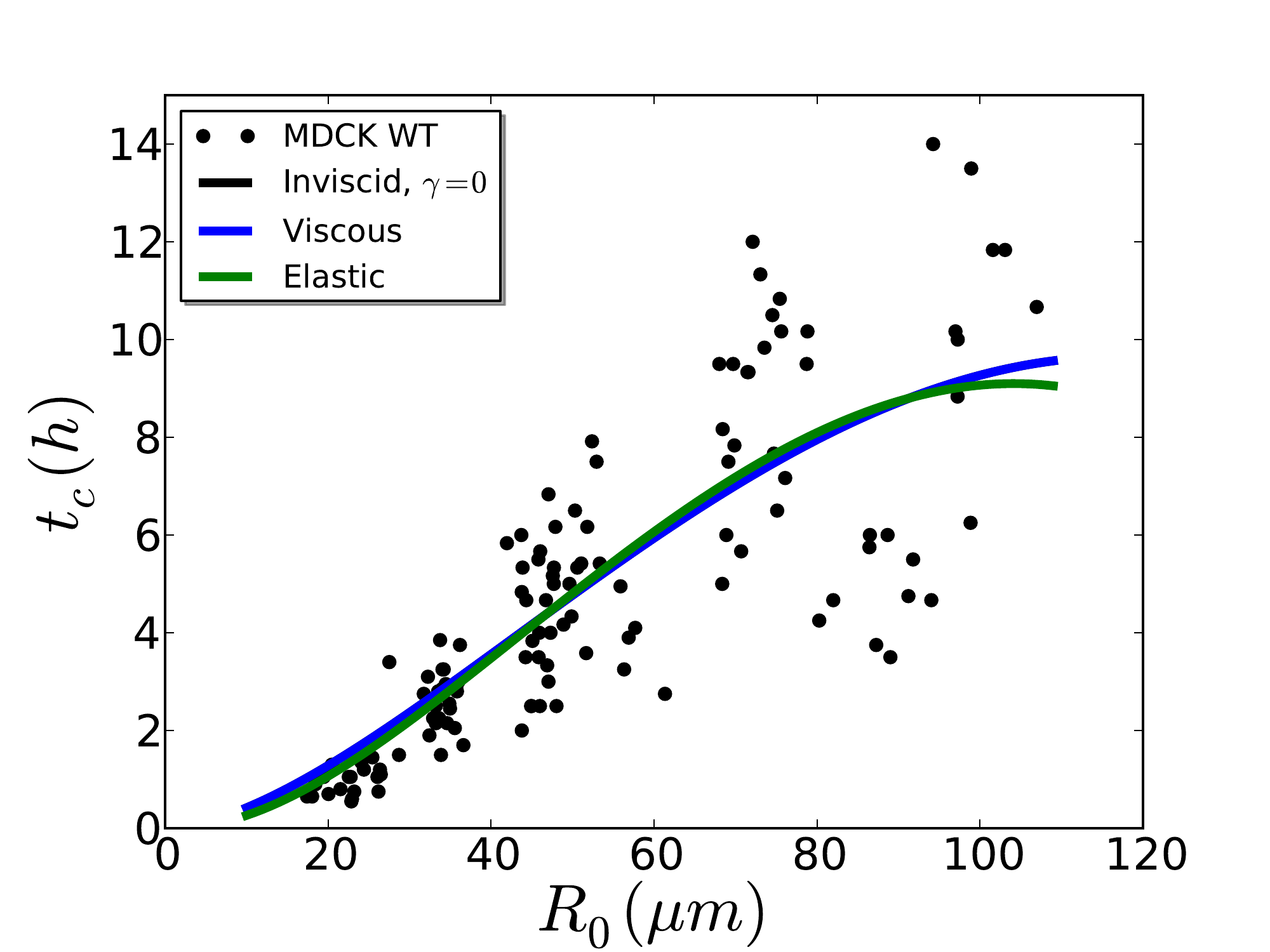}
\caption{\label{fig:tc:rheology} 
\textbf{Model selection:} MDCK wild type wounds.\\
\textbf{A Border forces.} The closure time 
$t_{\mathrm c}$ is plotted as a function 
of the initial effective radius $R_0$ (black circles), and fitted by
analytical expressions obtained when the epithelium is modeled
as an inviscid fluid:\\
- Equation~(\ref{eq:tc:R0:inviscid:nocable}): black line, 
$\sigma_{\rm p} \neq 0, \gamma  = 0$, constraints $D, R_{\mathrm{max}} \ge 0$,
fitted parameter values
$\sigma_{\rm p}/\xi = 353 \pm 38 \, \mu \mathrm{m}^2 \, \mathrm{h}^{-1}$,
$R_{\mathrm{max}} = 117 \pm 11 \, \mu \mathrm{m}$;\\
- Equation~(\ref{eq:tc:R0:inviscid:cable}): 
red line, $\sigma_{\rm p} \neq 0, \gamma \neq 0$, 
constraints $D, R_{\mathrm{max}}, R_{\gamma} \ge 0$,
fitted parameter values
$\sigma_{\rm p}/\xi = 247 \pm 108 \, \mu \mathrm{m}^2 \, \mathrm{h}^{-1}$,
$R_{\mathrm{max}} = 104 \pm 13 \, \mu \mathrm{m}$,
$R_{\gamma} = 7 \pm 11 \, \mu \mathrm{m}$;\\
- Equation~(\ref{eq:tc:R0:inviscid:nosigma}): 
blue line , $\sigma_{\rm p} = 0, \gamma \neq 0$, 
constraints $\gamma/\xi, R_{\mathrm{max}} \ge 0$,
fitted parameter values
$\gamma/\xi = 8592 \pm 606 \, \mu \mathrm{m}^3 \, \mathrm{h}^{-1}$,
$R_{\mathrm{max}} = 89 \pm 2 \, \mu \mathrm{m}$.\\
\textbf{B Tissue rheology.} The closure time $t_{\mathrm c}$ is plotted as 
a function of the initial effective radius $R_0$ (black circles), and
fitted by analytical expressions obtained when both lamellipodial protrusions 
and an actomyosin cable are taken into account 
($\sigma_{\rm p} \neq 0, \gamma \neq 0$):\\
- Equation~(\ref{eq:tc:R0:inviscid:nocable}):
black  line, inviscid fluid as in A;\\ 
- Equation~(\ref{eq:tc:R0:viscous:cable}): 
blue line, viscous fluid, 
constraints $D, R_{\mathrm{max}}, R_{\gamma}, R_{\eta} \ge 0$,
the (blue) fitted curve cannot be distinguished from the black curve, with 
identical parameter values of $D$ and $R_{\mathrm{max}}$,
and $R_{\gamma} = R_{\eta} = 0$.\\
- Equation~(\ref{eq:tc:R0:elastic:cable}): 
green line, elastic solid,
constraints $D, R_{\mathrm{max}}, \mu, \gamma \ge 0$,
the fit yields 
$\sigma_{\rm p}/\xi \approx  247 \, \mu \mathrm{m}^2 \, \mathrm{h}^{-1}$,
$R_{\mathrm{max}} = 104 \, \mu \mathrm{m}$,
$\frac{2 \mu}{\sigma_{\rm p}} = 0$, 
$R_{\gamma} \approx 7 \, \mu \mathrm{m}$, from which
we deduce $R_{\mathrm{e}} = - R_{\gamma} < 0$.
}
\end{figure}

First, we investigate whether cable tension 
may significantly contribute to force production at the margin
(Fig.~\ref{fig:tc:rheology}A).
Fitting closure time data 
with expression~(\ref{eq:tc:R0:inviscid:cable}),
obtained for an inviscid epithelium with a cable, we find that:
\begin{itemize}
\item[-]{values of $D$ and $R_{\mathrm{max}}$ are consistent within error bars
with those obtained without a cable;}
\item[-]{the length scale $R_{\gamma} = 7 \pm 11 \, \mu \mathrm{m}$ is small
    compared to $R_0$ ($R_{\gamma} \ll R_0$), as well as to the wound radius
($R_{\gamma} \ll R(t)$)
except in the late stages of closure \cite{Anon2012}.}
\end{itemize}
When the epithelium is modeled as an inviscid fluid, we conclude that 
the contribution of the actomyosin cable to the stress boundary condition
is negligible.
For the sake of completeness, we investigate the case where protrusive forces
are small compared to the cable tension ($R_{\gamma} \gg R_0$).
In this case, the closure time is given by
\begin{equation}
  \label{eq:tc:R0:inviscid:nosigma}
9 \frac{\gamma}{\xi} \,  t_{\mathrm c}(R_0) = R_0^3 \, \left( 1 + 
3 \ln \frac{R_{\mathrm{max}}}{R_0}  \right),
\end{equation}
which follows from integration of Eq.~\eqref{eq:R_dynamics_invisc} with
$\sigma_{\rm p} = 0$. This expression fits the closure time data 
rather poorly (Fig.~\ref{fig:tc:rheology}A):
protrusive forces at the margin cannot be neglected.

Second, we ask whether neglecting viscous stresses in the epithelium
is legitimate, and fit data with Eq.~(\ref{eq:tc:R0:viscous:cable})
(see Fig.~\ref{fig:tc:rheology}B). We obtain:
\begin{itemize}
\item[-]{values of $D$ and $R_{\mathrm{max}}$ consistent within error bars with
those found in the inviscid case without a cable;}
\item[-]{a length scale $R_{\gamma} = 7 \pm 51 \, \mu \mathrm{m}$,
consistent with a zero value;}
\item[-]{a viscous length scale $R_{\eta} = 0.01 \pm 8000 \, \mu \mathrm{m}$,
consistent with a zero value.}
\end{itemize}
We conclude that the actomyosin cable can be neglected in this case as well
($R_{\gamma} \ll R_0$), and that dissipation is dominated by friction 
with the substrate ($R_{\eta} \ll R_0$): epithelial viscosity
can be neglected.

Finally, we study closure time data taking into account elastic stresses,
and fit data with Equation~(\ref{eq:tc:R0:elastic:cable}), constraining the
parameters $D_s$, $R_{\mathrm{max}}$, $2 \mu/\sigma_{\rm p}$, and $R_{\gamma}$ 
to be positive. The fitted value of $2 \mu/\sigma_{\rm p}$ is consistent with 
zero: elastic forces are vanishingly small when compared
to protrusive forces. In addition, the fitted values of 
$D_s = D$, $R_{\mathrm{max}}$ and $R_{\gamma}$ are consistent with those 
obtained for an inviscid fluid when the cable line tension is
taken into account.
In this case, Equation~(\ref{eq:tc:R0:elastic:cable})
reduces to Equation~(\ref{eq:tc:R0:inviscid:cable}).

Altogether, we find that the model of the monolayer as an
inviscid fluid describes wild-type MDCK data 
satisfactorily, and that viscous and elastic contributions to the stress 
are negligibly small.
Furthermore, the contribution of the cable to force production
is small compared to that of lamellipodia. We hypothesize that 
the main function of the contractile circumferential cable 
is to stabilize the free epithelial boundary.
Since $R_{\gamma} \ll R_0$ in all cases considered, we neglect 
cable tension in the following and set $\gamma = 0$ unless explicitly 
specified otherwise.

\begin{figure}[!t]
\textbf{A}
\includegraphics[width=7.7cm]{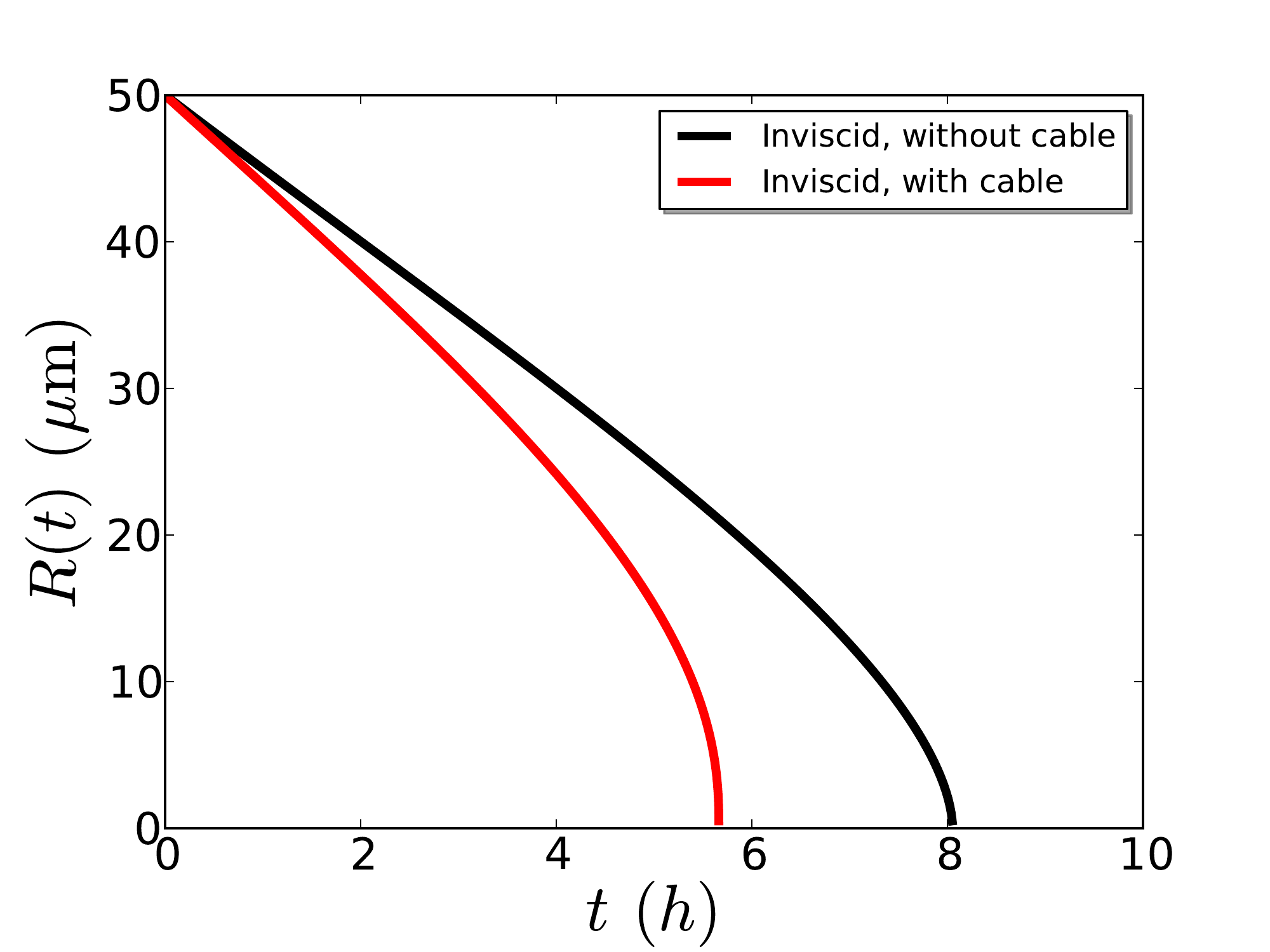}
\textbf{B}
\includegraphics[width=7.7cm]{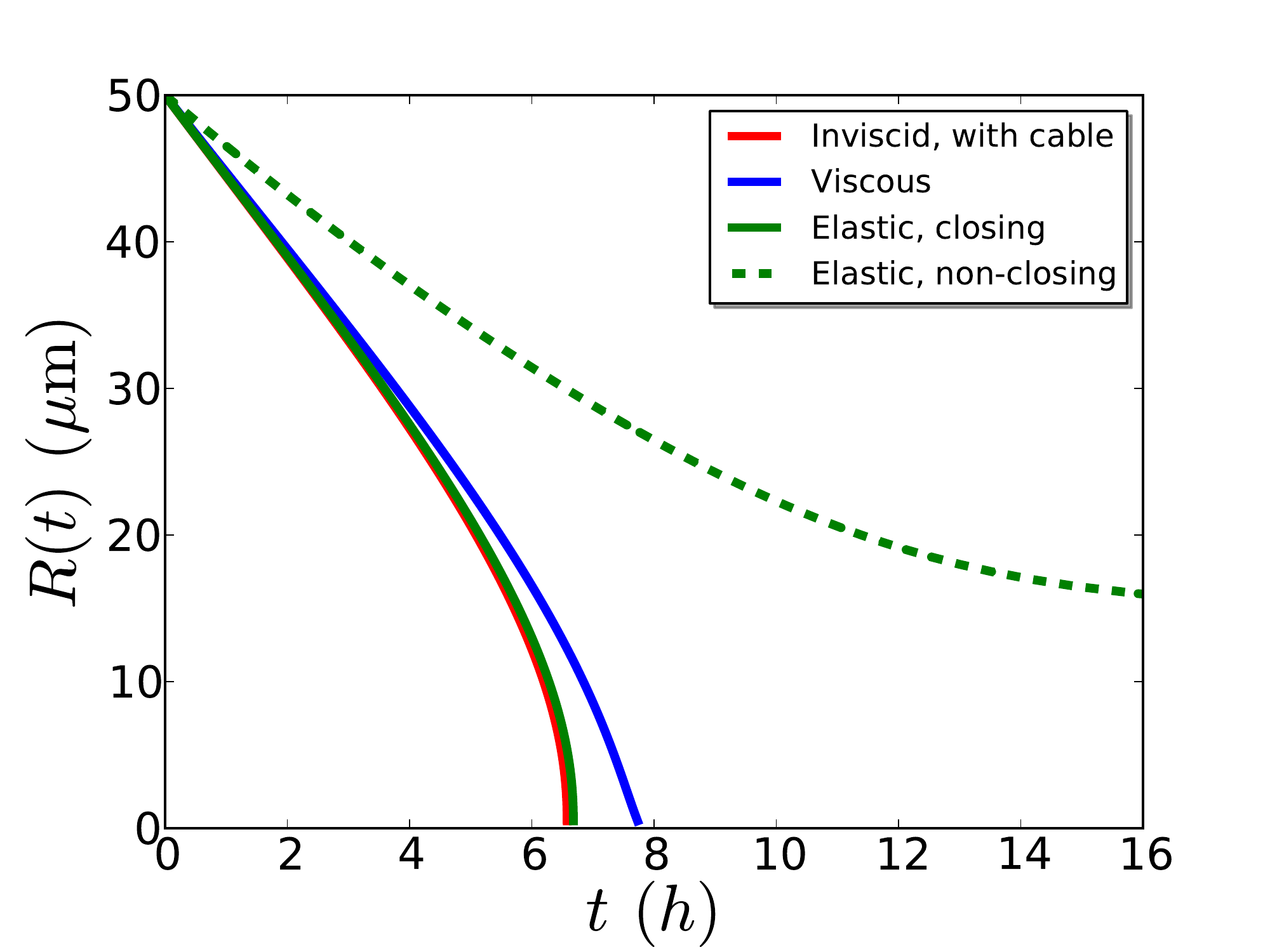}
\caption{\label{fig:model:traj} 
\textbf{Model predictions.}\\
Plots of individual trajectories of the wound radius $R(t)$,
$R_0 = 50 \, \mu \mathrm{m}$.\\
\textbf{A:} Contribution of border forces for an inviscid fluid. 
Plots of $R(t)$ as given by Eq.~(\ref{eq:inviscid:t:R}),
$D = 200 \, \mu \mathrm{m}^2 \, \mathrm{h}^{-1}$,
$R_{\mathrm{max}} = 110 \, \mu \mathrm{m}$ (black curve, without cable);
and by Eq.~(\ref{eq:ttilde_invisc}) same values of $D$ and 
$R_{\mathrm{max}}$, $R_{\gamma} = 10 \, \mu \mathrm{m}$
(red curve, with cable).\\
\textbf{B:} Rheology. Plots of $R(t)$ as given by 
Eq.~(\ref{eq:ttilde_invisc}) (red curve, inviscid liquid, same as in a);
Eqs.~(\ref{eq:viscous:t:R:1}-\ref{eq:viscous:t:R:2}), with 
$R_{\eta} = 10 \, \mu \mathrm{m}$
(blue curve, viscous liquid); 
Eqs.~(\ref{eq:solution_incompressible_elastic_0}-\ref{eq:solution_incompressible_elastic}), with 
$\frac{2 \mu}{\sigma_{\rm p}} = 0.1$
(solid green curve, elastic solid, closing);
Eq.~(\ref{solution_incompressible_elastic_positive}), with 
$R_{\mathrm{e}} = 15 \, \mu \mathrm{m}$
(dashed green curve, elastic solid, non-closing);
The values of $D$, $R_{\mathrm{max}}$ and  $R_{\gamma}$ are the same as in A.
}
\end{figure}

\subsection{Closure trajectories}
\label{sec:data:traj}

In sections~\ref{sec:data:traj:closing} and \ref{sec:data:traj:nonclosing},
we examine the individual trajectories of closing and 
non-closing wounds.

\begin{figure}[!t]
\textbf{A}
\includegraphics[width=7.7cm]{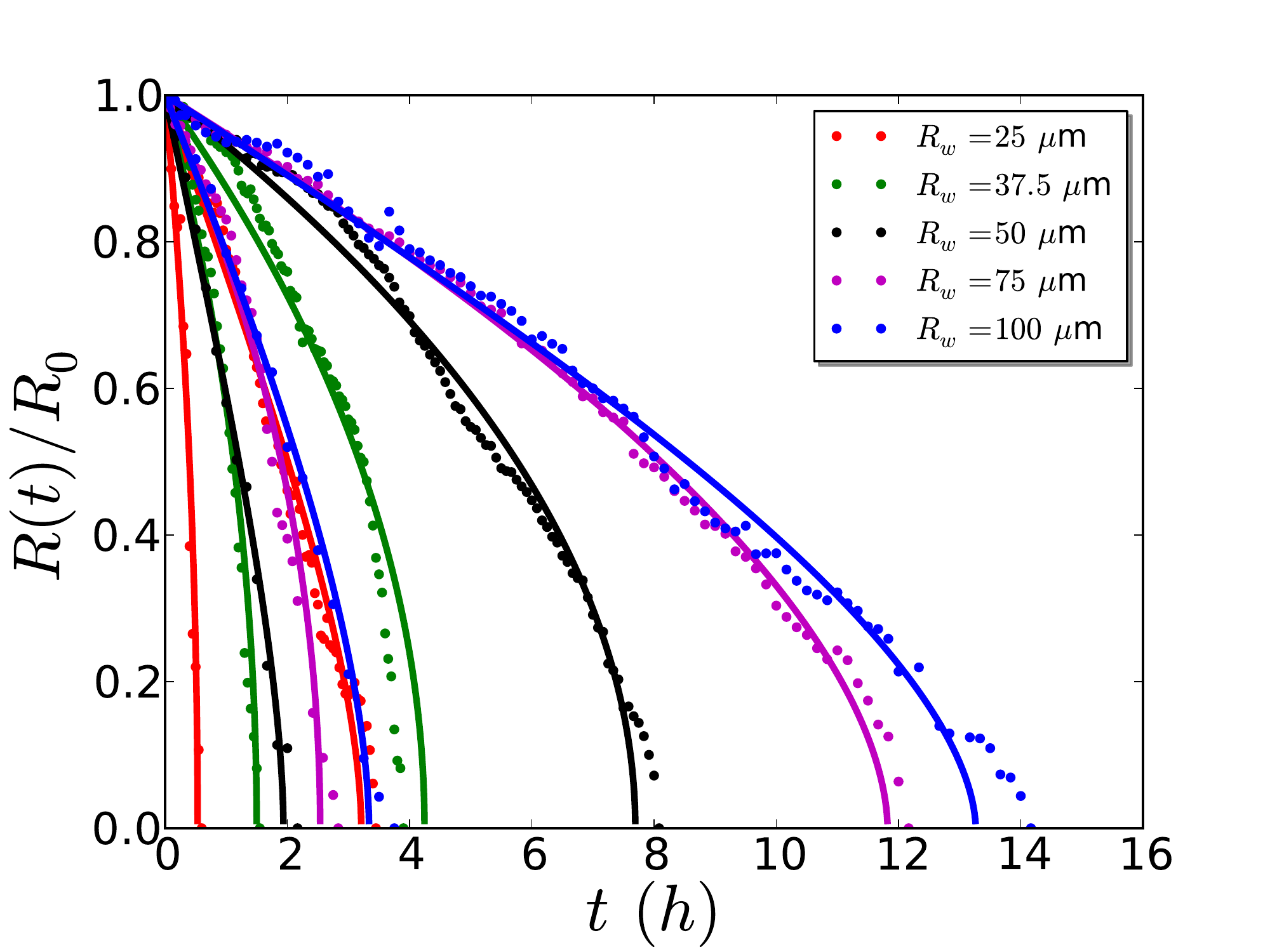}
\textbf{B}
\includegraphics[width=7.7cm]{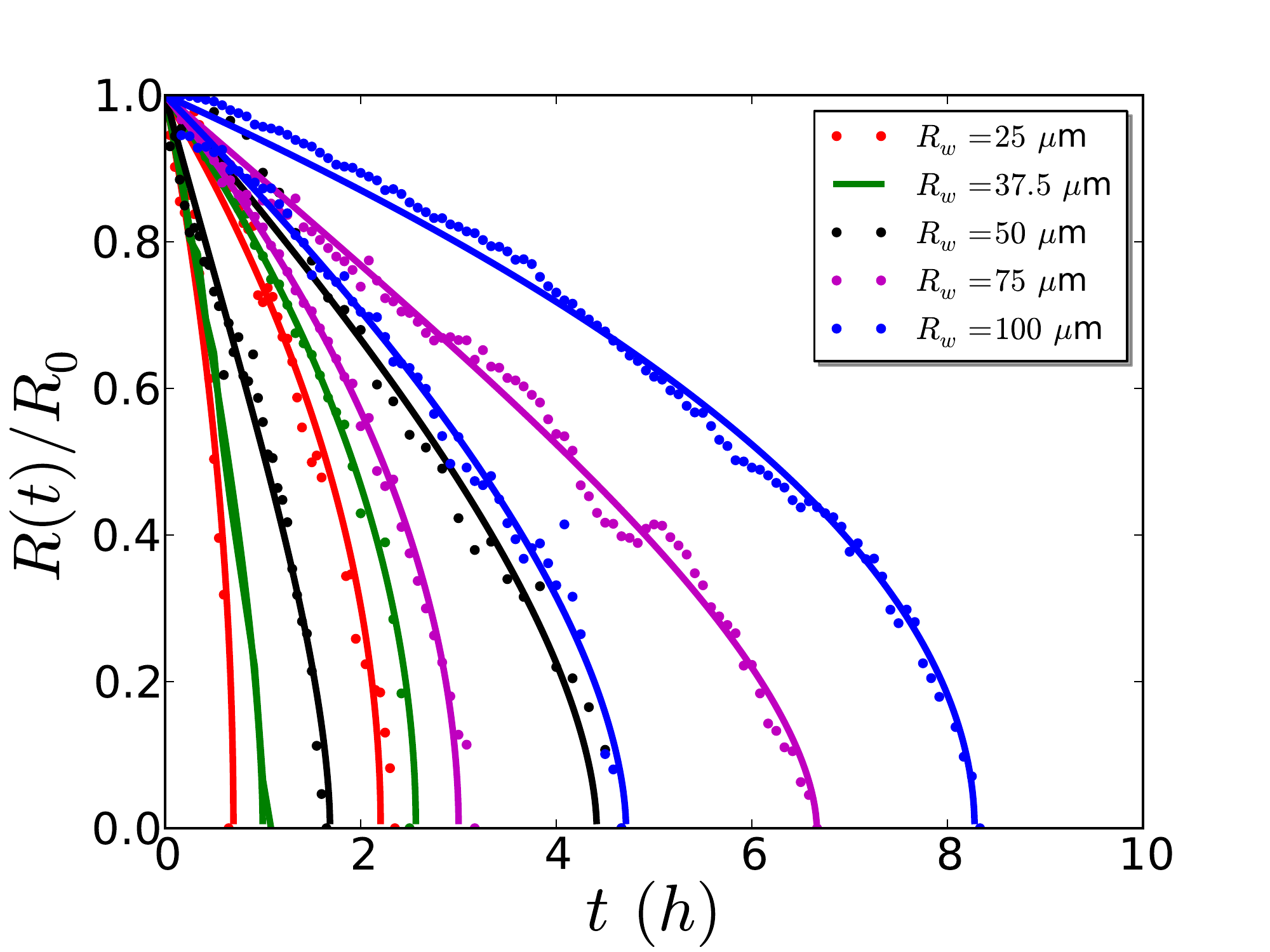}
\textbf{C}
\includegraphics[width=7.7cm]{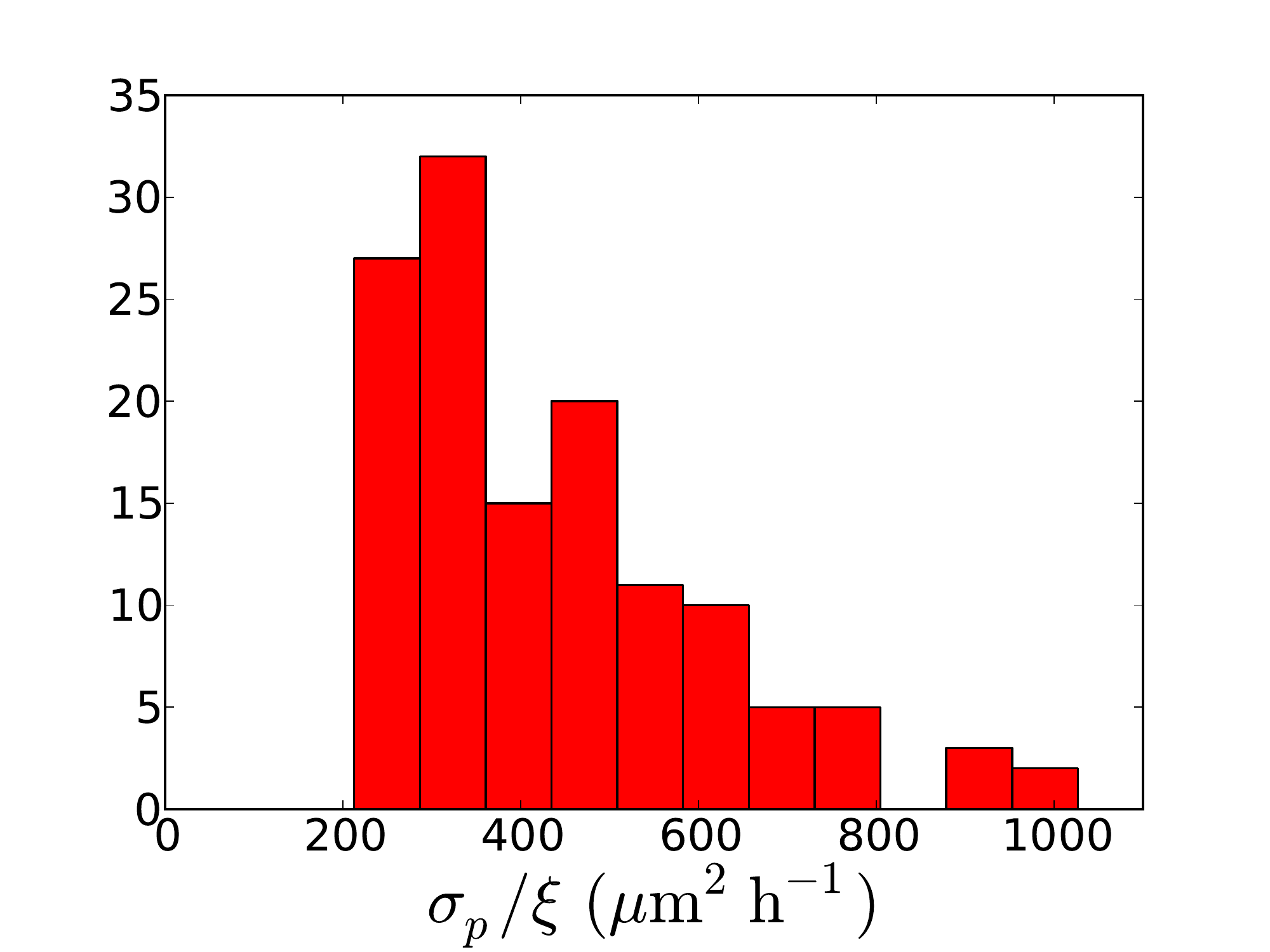}
 \textbf{D}
\includegraphics[width=7.7cm]{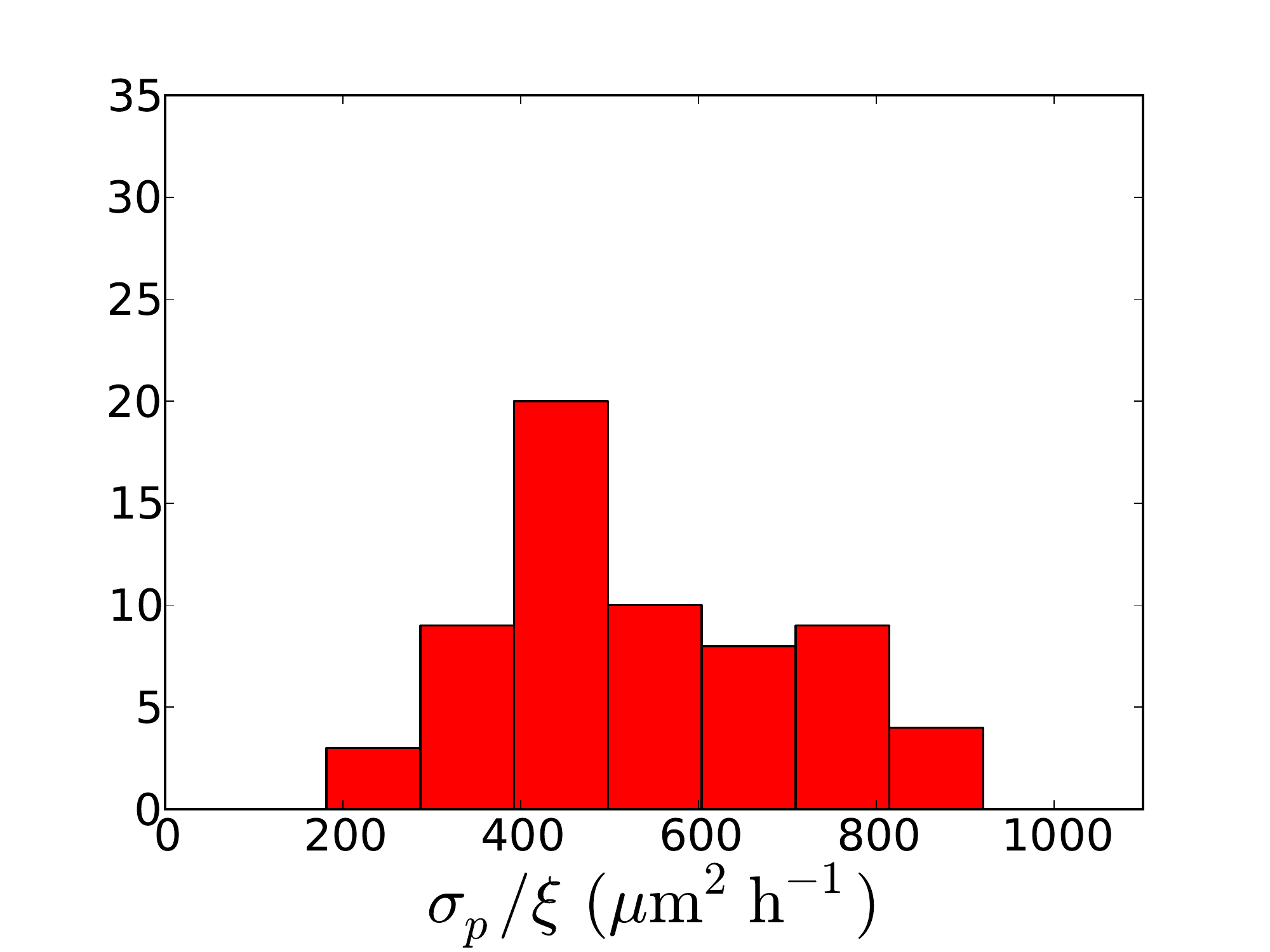}
\caption{\label{fig:hist:closing} 
\textbf{Trajectories $R(t)$ of closing wounds.}
\textbf{A, C:} MDCK wild type wounds;
\textbf{B, D:} HEK-HT wounds.\\ 
\textbf{A, B:} Fit of trajectories $R(t)$ with 
Equation~(\ref{eq:inviscid:t:R}). For clarity, we show only 
two trajectories (circles) and their fits (solid curves) 
per pillar size $R_{\mathrm{w}}$, corresponding to 
the shortest and longest closure time observed at a given $R_{\mathrm{w}}$.
The normalized effective radius $R(t)/R_0$ is plotted as a function of time $t$.
\\
\textbf{C, D:} Histogram of estimates of
the epithelization coefficient (see text for details).\\
}
\end{figure}

\subsubsection{Closing wounds}
\label{sec:data:traj:closing}

For brevity, we focus on MDCK-WT and HEK-HT wounds, and fit 
Equation~(\ref{eq:inviscid:t:R}) to data,
using for convenience time as a function of radius $t(R)$.
In Section~\ref{sec:data:tc}, we showed that the simplest model of the
monolayer as an inviscid fluid driven by cell protrusions at the margin
suffices to describe closure time data. We therefore fit 
trajectories using the same model (see Fig.~3a), 
obtain one set of physical parameters per wound,
and check the consistency of our results.

Since $R_{\mathrm{max}}$ was previously found to vary little, 
we constrain $R_{\mathrm{max}}$ to belong to the $95 \%$ confidence
interval obtained from closure time data 
(see the caption of Fig.~\ref{fig:tc:inviscid} for numerical values).
The distributions of epithelization coefficients obtained by 
fitting Equation~(\ref{eq:inviscid:t:R}) to data are shown in 
Fig.~\ref{fig:hist:closing}, for MDCK-WT and HEK-HT wounds, 
with mean values $\pm$ standard deviations given by:
\begin{itemize}
\item[-]{MDCK wild type wounds:
$\sigma_{\rm p}/\xi =  424 \pm  170 \, \mu \mathrm{m}^2 \, \mathrm{h}^{-1}$;} 
\item[-]{HEK-HT wounds:
$\sigma_{\rm p}/\xi =  522 \pm  165 \, \mu \mathrm{m}^2 \, \mathrm{h}^{-1}$.} 
\end{itemize}

For both cell types, the confidence intervals obtained from fitting
closure time data belong to the above intervals: 
the two measurement methods are consistent. Trajectories are noisy, due to
intrinsic variabity, but also to possible pixelization errors
when determining the area of the cell-free domain. Fitting 
individual trajectories leads to a higher dispersion of estimated 
parameter values. We therefore prefer to use closure time data 
for parameter estimation whenever closure is complete.

\subsubsection{Non-closing wounds}
\label{sec:data:traj:nonclosing}
 
\begin{figure}[!t]
\textbf{A}
\includegraphics[width=7.7cm]{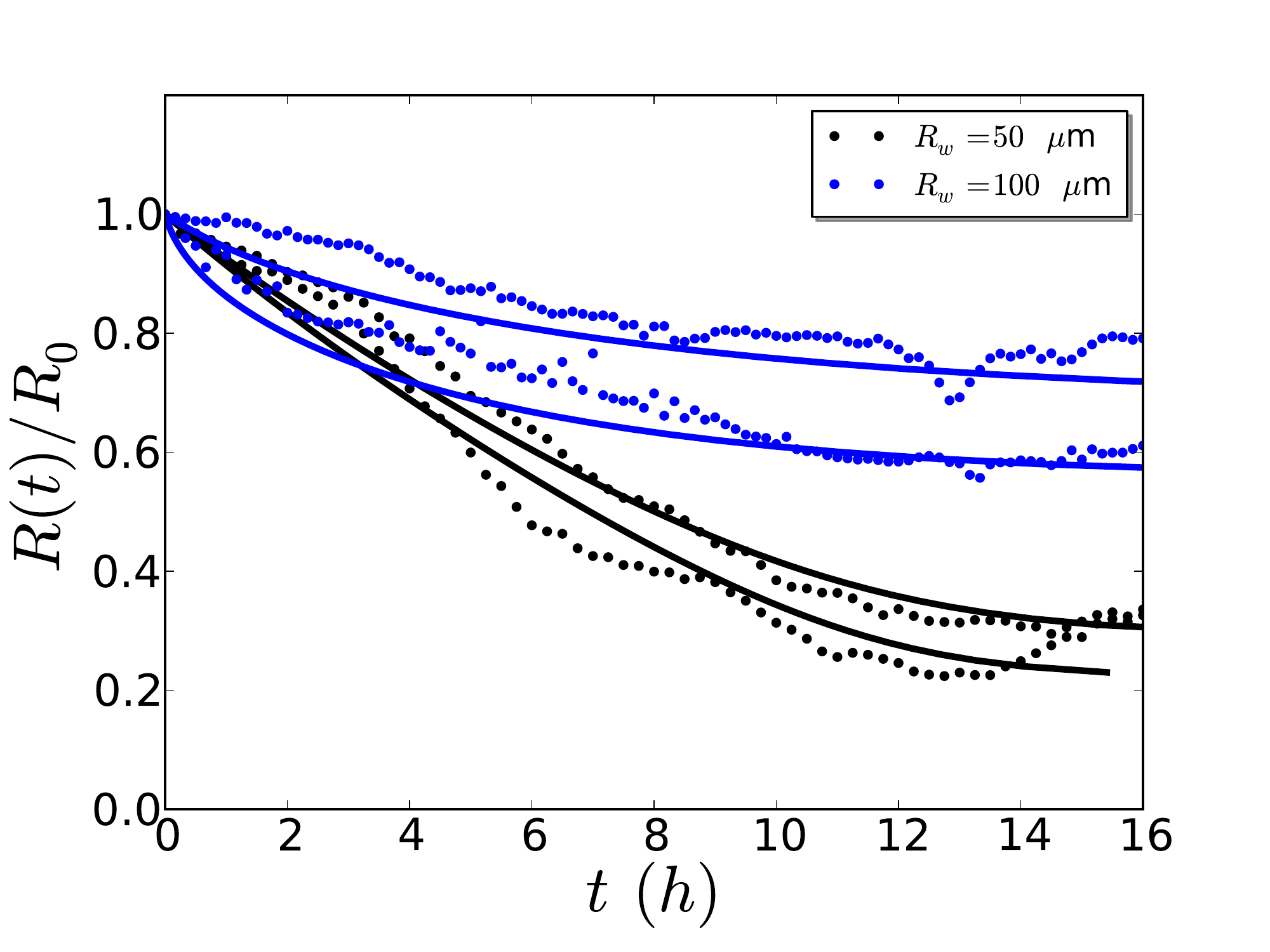}
\textbf{B}
\includegraphics[width=7.7cm]{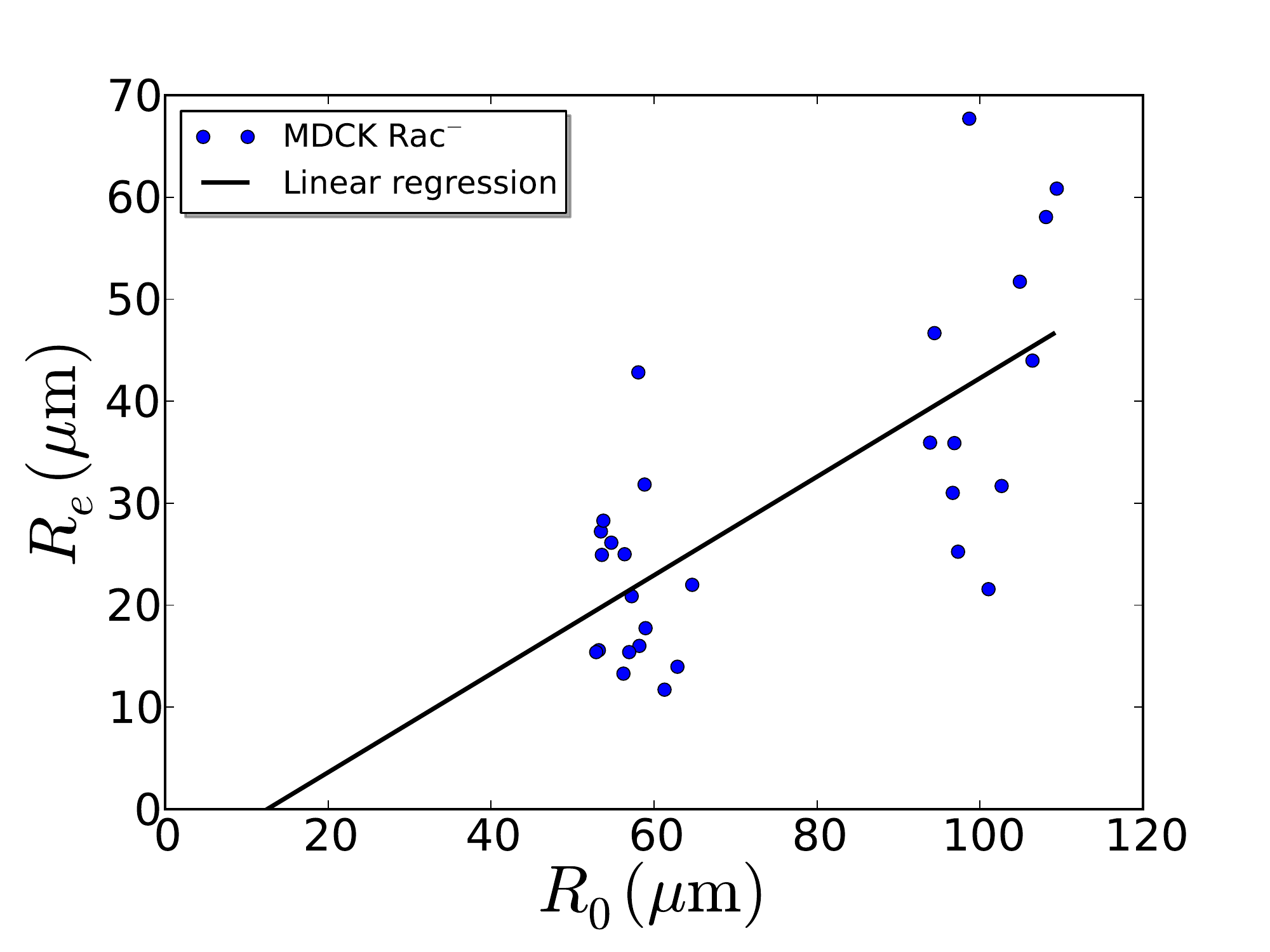}
\textbf{C}
\includegraphics[width=7.7cm]{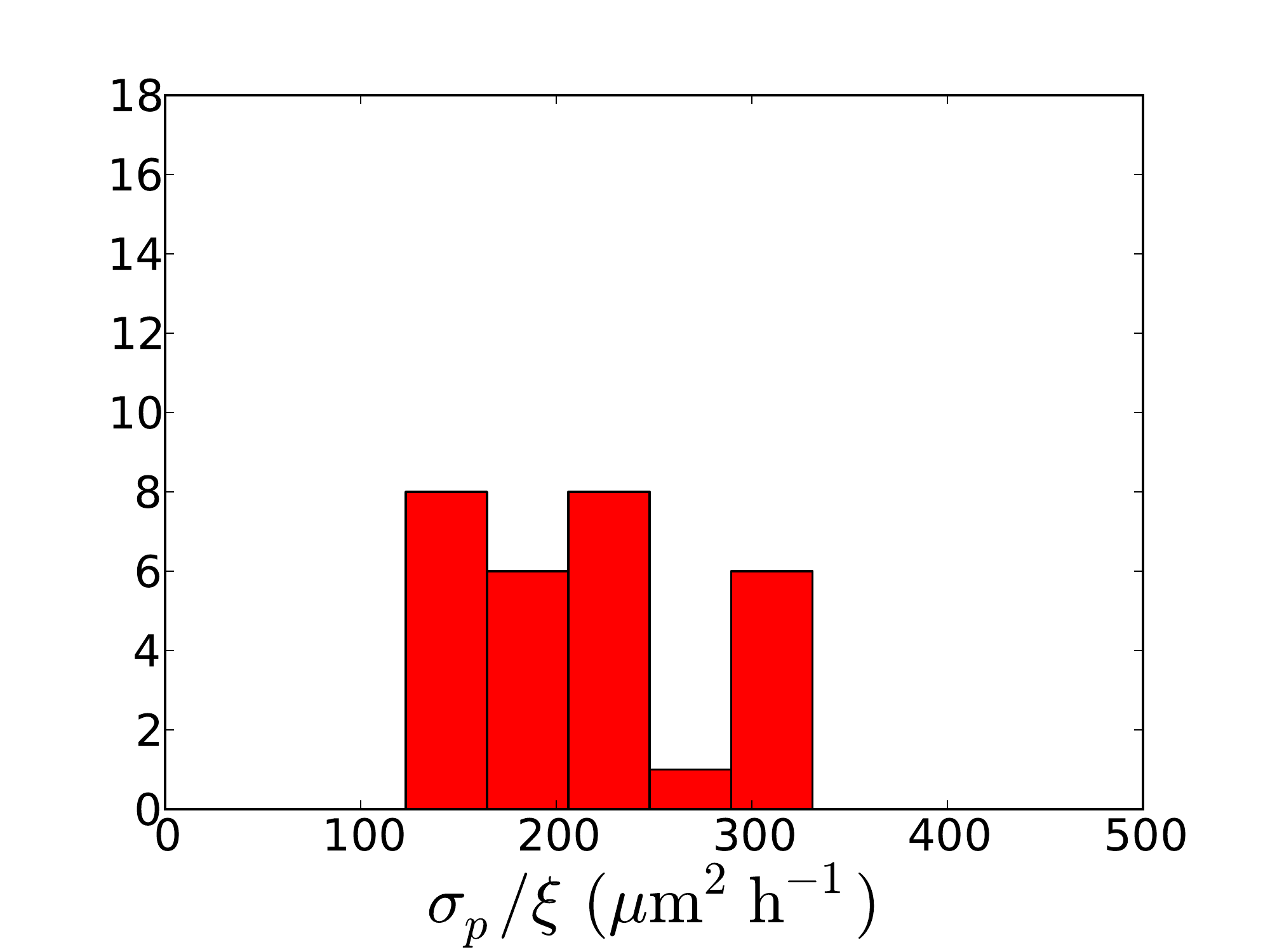}
\textbf{D}
\includegraphics[width=7.7cm]{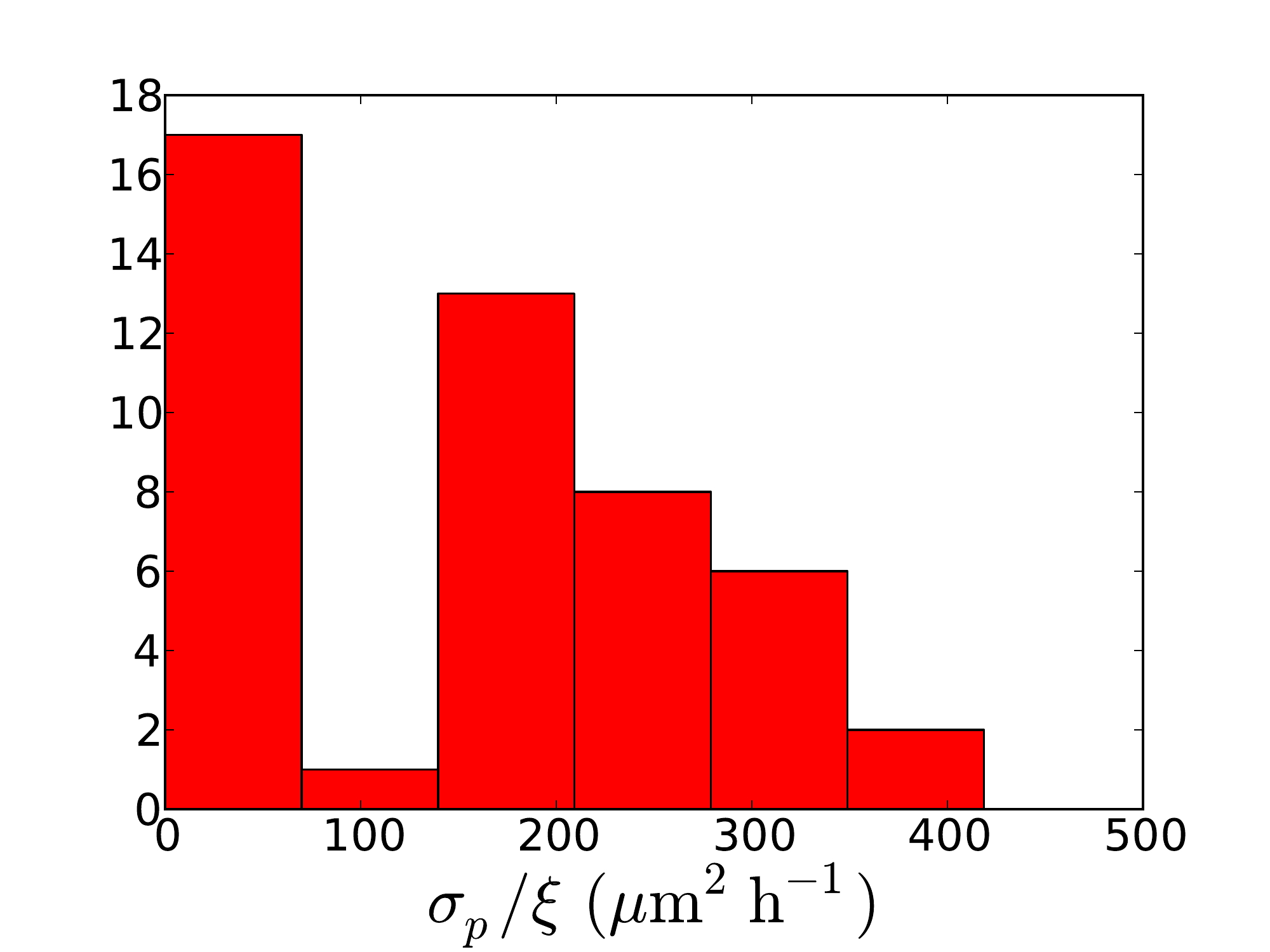}
\caption{\label{fig:hist:nonclosing} 
\textbf{MDCK Rac$^-$ assay}\\
\textbf{A-C: Non-closing wounds}\\
\textbf{A: Trajectories.}
Normalized effective radius $R(t)/R_0$ as a function of time $t$.
For illustrative purposes, we show only two trajectories $t(R)$ 
per pillar size $R_{\mathrm{w}}$ (solid curves) and their fit by
Equation~(\ref{solution_incompressible_elastic_positive})
(dashed curves), with the constraints 
$D_S \ge 0$, $R_{\mathrm{max}} \in [96 \; 114] \, \mu \mathrm{m}$
(confidence interval obtained from closure time data), 
$R_{\mathrm e} = \mathrm{min} \, R(t)$ 
(Equation~(\ref{solution_incompressible_elastic_positive}) 
is defined only for $R > R_{\mathrm e}$).
 \\
\textbf{B: Equilibrium Radius} $R_{\mathrm e}$ 
(estimated as $R_{\mathrm e} = \mathrm{min} \, R(t)$) vs. initial radius $R_0$.
The linear regression line (black solid line, $R_{\mathrm e} = a R_0 + b$) 
has coefficients $a = 0.5 \pm 0.1$, $b = -6 \pm 7 \, \mu$m.
\\
\textbf{C: Histogram of parameter estimates.} 
The epithelization coefficient is estimated as  $D = D_s/2$ 
(from $2 \mu/\sigma_{\rm p} \approx 1$), where $D_s$ is obtained 
by nonlinear curve fitting of the trajectory, as in (a).
\\
\textbf{D: Closing wounds.} Histogram of the epithelization coefficient.
Closing trajectories are fitted as in Fig.~3b.
}
\end{figure}

We finally turn to the non-closing wounds observed in MDCK Rac$^-$ assays. 
Among the models presented in Section~\ref{sec:model}, the only case
where the final radius is strictly positive is that of an elastic
epithelium with $R_{\mathrm e} > 0$, or $2 \mu > \gamma/R_0$.
In  Fig.~\ref{fig:hist:nonclosing}A, we show that
individual trajectories are fitted satisfactorily by
Equation~(\ref{solution_incompressible_elastic_positive}).
The equilibrium radius $R_{\mathrm e}$ increases with the initial effective 
radius $R_0$ (Fig.~\ref{fig:hist:nonclosing}B), 
as predicted by Equation~(\ref{def:Re}). A linear regression of $R_{\mathrm e}$ 
vs. $R_0$ yields the estimates 
\begin{eqnarray}
  \label{eq:elast:a}
  \frac{2 \mu}{2 \mu + \sigma_{\rm p}} &=& 0.5 \pm 0.1\\
  \label{eq:elast:b}
  \frac{\gamma}{2 \mu + \sigma_{\rm p}} &=& 6 \pm 7 \; \mu\mathrm{m}.
\end{eqnarray}
From (\ref{eq:elast:a}), we deduce that 
$\mu/\sigma_{\rm p} \approx 0.5$. Assuming that
the Rac pathway has a limited influence on the epithelial elasticity,
this suggests that Rac  inhibition leads to lower values of the 
protrusive stress (compared to wild type assays), 
of the order of the elastic modulus.
Since $\mu/\sigma_{\rm p} \approx 0.5$, 
Equation~(\ref{eq:elast:b})
yields $R_{\gamma} \approx 10 \, \mu$m: the actomyosin cable contributes
significantly to force production in non-closing
Rac$^-$ assays when, \emph{e.g.}, $R_{\mathrm{w}} = 50 \, \mu$m.

Fitting non-closing trajectories, we obtain  estimates of the coefficient
$D_s = D \left( 1 + \frac{2 \mu}{\sigma_{\rm p}} \right)$. Using 
$\mu/\sigma_{\rm p} \approx 0.5$, we expect that $D \approx 0.5 \, D_s$.
In Fig.~\ref{fig:hist:nonclosing}C, we plot the
histogram of epithelization coefficients defined for simplicity
as  $D = 0.5 \, D_s$. We find
$\sigma_{\rm p}/\xi =  180 \pm 45 \, \mu \mathrm{m}^2 \, \mathrm{h}^{-1}$
(mean value $\pm$ standard deviation, $N = 29$).
Fitting Rac$^-$ closing trajectories with 
Equation~(\ref{eq:inviscid:t:R}) for an inviscid 
epithelium, we obtain 
$\sigma_{\rm p}/\xi =  230 \pm 66 \, \mu \mathrm{m}^2 \, \mathrm{h}^{-1}$ 
($N = 30$), a value slightly higher than the previous estimate 
obtained for non-closing wounds. Note that both estimates are consistent with
that obtained from fitting time closure data.

A balance between driving forces
at the margin and a bulk elastic restoring force 
explains the positive value of the equilibrium radius
observed in these assays. 
A word of caution seems however in order.
Although $R(t)$ plateaus on a time scale of the order of $15$~h, one cannot
exclude that a ``non-closing wound'' may in fact heal completely on a 
time scale much longer than the available observation time, over which
cell divisions may become relevant and need to be taken into account.

\subsection{Physical parameters of epithelization}
\label{sec:data:param}

The epithelization coefficient $D = \sigma_{\rm p}/\xi$ estimated for 
wild type MDCK wounds was of the order of 
$350 \, \mu \mathrm{m}^2 \, \mathrm{h}^{-1}$,
or $10^{-1} \, \mu \mathrm{m}^2 \, \mathrm{s}^{-1}$.
Using the order of magnitude of cell protrusive forces 
$F_p \approx 1$ nN \cite{Prass2006}, the two-dimensional
protrusive stress $\sigma_{\rm p}$ is of the order of
$F_p/L$, where $L$ is the typical lateral extension of a cell.
Using $L \approx 10 \, \mu \mathrm{m}$, we find
$\sigma_{\rm p} \approx 10^{-1} \, \mathrm{nN}\,\mu \mathrm{m}^{-1}$.
We then deduce the order of magnitude of the friction coefficient
$\xi \approx 1 \, \mathrm{nN} \, \mu\mathrm{m}^{-3} \, \mathrm{s}$,
here for a cell monolayer on a glass substrate.
Interestingly, this value is consistent with that proposed 
in \cite{Arciero2011}, using very different assumptions 
to model epithelization.

Compared to wild type MDCK assays, the epithelization coefficient $D$ 
adopted a lower value under Rho inihibition, and was further reduced 
by Rac inhibition. A lower value of the ratio  $\sigma_{\rm p}/\xi$ 
corresponds to a lower value of $\sigma_{\rm p}$ and/or to a higher value of $\xi$.
In the case of Rac$^-$ assays, it is now well established that Rac is 
responsible, through the activation of the Arp2/3 complex, 
for actin polymerization at the leading edge of a migrating cell 
\cite{Hall1998,Jaffe2005,Etienne-Manneville2002}, which is necessary 
for force production by lamellipodia. 
The lower value of $D$ in Rac$^-$ assays may well be explained 
by this effect only. However, Rac inhibition may also modify 
the value of $\xi$: indeed the Rac pathway  
is also known to be involved in the formation of focal contacts 
(see, \emph{e.g.}, \cite{Rottner1999}).

On general physical grounds \cite{Schallamach1963,Gerbal2000}, a simple
expression for the friction coefficient is given by $\xi = n k \tau$,
where $n$, $k$ and $\tau$ respectively denote the average density
of adhesive bonds, the bond spring constant, and the average binding 
time. These three quantities are related to the formation of adhesive bonds, 
to their maturation state and to their turnover. 
The influence of the Rho and Rac GTPases on these three 
mutally interacting biological processes
is complex, often with antagonistic effects on any two of them
\cite{Hall1998,Rottner1999,Danen2005,Etienne-Manneville2002,Jaffe2005}.
On the basis of current knowledge, predicting the effect of Rho and Rac 
inhibition  on epithelium-substrate friction seems very difficult,
all the more so since conclusions drawn from single-cell motility assays may 
not carry over to the case of collective migration of a cell monolayer.
Still, it has been shown that Rho is not implicated in the polarization 
of actin at the leading edge of a migrating cell and that its inhibition 
can even enhance motility in certain cell types \cite{Nobes1999}. 
We conjecture that the lower value of $D$ in Rho$^-$ assays may be due
to a higher value of $\xi$.
This may be explained by the implication of Rho in regulating the 
turn-over of adhesion complexes, more stable under Rho inhibition,
thus leading to a higher $\tau$, and possibly to a higher $\xi$ 
\cite{Danen2005}.  However, existing data regarding the effect of 
Rho on $n$ and $k$ is inconclusive: it has for instance been observed 
that Rho$^-$ assays lead to a lower integrin density \cite{Ballestrem2001}. 
Our measurement may be seen as direct evidence for the effect of Rho 
inhibition on the epithelium-substrate friction coefficient,
and may  be used as a basis towards a better understanding 
of the role plaid by the Rho GTPase in regulating the formation, 
the maturation, and the turn-over of cell-substrate adhesive bonds
in epithelia.

Fits of closing and non-closing trajectories in Rac$^-$ assays showed 
that the epithelization coefficient was larger when closure is complete.
Neglecting the cable line tension $\gamma = 0$, the equilibrium radius
reads $R_{\rm e} = R_0/(1 + 2\mu/\sigma_{\rm p}) \ge 0$.
Our model suggests that closure is incomplete as soon as 
$R_{\mathrm e} > a$, where $a$ is the cellular length scale below which 
micro-scale mechanisms operate to terminate epithelization.
For simplicity, we ignore the possible influence 
of Rac inhibition on the epithelial elastic modulus, through, {\it e.g.}
the dynamics and density of cell-cell adhesions \cite{Etienne-Manneville2002}.
The condition $R_{\mathrm e} > a$ corresponds to a threshold value 
$\sigma_{\rm p, c}$ of the protrusive stress,
$\sigma_{\rm p} < \sigma_{\rm p, c} = 2 \mu \, (R_0 - a)/a \simeq 2 \mu R_0/a$,
that increases with $R_0$. 
Given the observed experimental variability, we expect the value of 
$\sigma_{\rm p}$ to fluctuate from wound to wound in a given Rac$^-$
assay. For smaller wounds, crossing the threshold $\sigma_{\rm p, c}$ 
is less likely: indeed the fraction of non-closing wounds is an 
increasing function of initial radius (Fig.~2C).
Altogether, our analysis suggests that Rac inhibition lowers the ratio 
$\sigma_{\rm p}/\mu$ so that epithelial elasticity can no longer be 
neglected.


\newpage
\noindent
MOVIE 1. \textbf{Closure of a large MDCK circular wound.}\\
A MDCK-actin-GFP wound ($R_{\mathrm{w}} = 250 \, \mu$m) is imaged 
in epifluorescence for $14$ h. Scale bar: $200 \, \mu$m.
Three leader cells formed at the edge of the wound and then drove 
multicellular fingers hence deforming the initial circle 
(Fig.~\ref{fig:circ}). The fingers eventually met in the center 
and the leader cells switched back to a classical epithelial phenotype. 
The remaining secondary wounds then proceeded to heal in a much more regular 
fashion without showing any formation of leader cells (Movie 2).

\bigskip
\noindent
MOVIE 2. \textbf{Closure of ``secondary'' wounds.}\\
Close-up on the secondary wounds from the experiment seen in Movie 1,
imaged for $8.3$ h. Scale bar: $100 \, \mu$m. 
Neither leader cells and nor margin roughening are seen.

\bigskip
\noindent
MOVIES 3-5. \textbf{Closure of small circular wounds.}\\
Three examples of time lapse movies made in phase contrast microscopy 
showing the typical closure of a wound for, respectively:
\begin{itemize}
\item wild type MDCK cells, $R_{\mathrm{w}} = 50 \, \mu$m for $6$ h $30$; 
scale bar: $100 \, \mu$m;
\item HEK-HT cells, $R_{\mathrm{w}} = 50 \, \mu$m for $4$ h; 
scale bar: $100 \, \mu$m;
\item HEK-RasV12 cells, $R_{\mathrm{w}} = 75 \, \mu$m for $3$ h $30$; 
scale bar: $150 \, \mu$m.
\end{itemize}
Direct inspection shows that the protrusive activity is enhanced in 
the last case, with a closure time shorter compared to a 
smaller wild type HEK wound.

\bigskip
\noindent
MOVIE 6. \textbf{Dynamics of lamellipodial activity.}\\
A MDCK-LifeAct-GFP wound ($R_{\mathrm{w}} = 25 \, \mu$m) was imaged by confocal 
microscopy for $3$ h. Scale bar: $25 \, \mu$m. 
The optical slice was very close to the surface as this is the position 
where lamellipodia develop. For this reason, stress fibers were apparent 
but the membranes between cells were not. We observed a high number and 
a large activity of these lamellipodia that could be recognized as waves 
of actin in the bulk of the tissue. Of note, high laser power was needed 
to observe these lamellipodia and the dynamics of closure was drastically 
reduced in those experiments probably due to phototoxicity. 

\bigskip
\noindent
MOVIE 7. \textbf{Laser ablation of the entire cable.}\\
A MDCK LifeAct-GFP wound ($R_{\mathrm{w}} = 25 \, \mu$m) is imaged by 
confocal microscopy from $t=30$ min after removal of the pillars. 
The actin cable is then fully ablated and the retraction of the edge is 
imaged for $1$ min. Note the dynamic retraction of the edge of the wound. 
Scale bar : $10 \, \mu$m.

\bigskip
\noindent
MOVIE 8. \textbf{A non closing MDCK Rac$^-$ wound.}\\
A MDCK wound ($R_{\mathrm{w}} = 100 \, \mu$m) under Rac inhibition was imaged 
in phase contrast. The movie runs for $17.5$ h. Scale bar: $100 \, \mu$m.

\end{document}